\documentclass[%
 superscriptaddress,
 twocolumn,
 showpacs,
 preprintnumbers,
 amsmath,amssymb,
 prd,
floatfix,
]{revtex4-1}

\usepackage{graphicx}% Include figure files
\usepackage{dcolumn}% Align table columns on decimal point
\usepackage{bm}% bold math
\usepackage{mathrsfs}
\usepackage{url}
\usepackage{xcolor}
\usepackage{subfigure}
\usepackage{hyperref}% add hypertext capabilities
\usepackage[mathlines]{lineno}% Enable numbering of text and display math
\begin{document}

\preprint{FERMILAB-PUB-21-397-E}
% \preprint{arXiv:}

\title{Measurement of Charge and Light Yields for $^{127}$Xe $L$-Shell Electron \\Captures in Liquid Xenon}% Force line breaks with \\

\author{D. J. Temples}
 \email{dtemples@fnal.gov}
 \thanks{Formerly at 2.
    %Northwestern University, Department of Physics \& Astronomy, Evanston, IL  60208-3112, USA
 }
 \affiliation{%
 Fermi National Accelerator Lab (FNAL), Astrophysics Department, Batavia, IL 60510-5011, USA
}%
%  \affiliation{%
%  Northwestern University, Department of Physics \& Astronomy, Evanston, IL  60208-3112, USA
% }%

\author{J. McLaughlin}%
\affiliation{%
 Northwestern University, Department of Physics \& Astronomy, Evanston, IL  60208-3112, USA
}%

\author{J. Bargemann}
\affiliation{
University of California, Santa Barbara, Department of Physics, Santa Barbara, CA 93106-9530, USA
}

\author{D. Baxter}
\thanks{Formerly at 2.
    %Northwestern University, Department of Physics \& Astronomy, Evanston, IL  60208-3112, USA
}
\affiliation{%
 Fermi National Accelerator Lab (FNAL), Astrophysics Department, Batavia, IL 60510-5011, USA
}%

\author{\\A. Cottle}
\thanks{Formerly at 1.
    %Fermi National Accelerator Laboratory (FNAL), Astrophysics Department, Batavia, IL 60510-5011, USA
}
\affiliation{%
 University of Oxford, Department of Physics, Oxford OX1 3RH, UK
}%

\author{C. E. Dahl}
\affiliation{%
 Fermi National Accelerator Lab (FNAL), Astrophysics Department, Batavia, IL 60510-5011, USA
}%
\affiliation{%
 Northwestern University, Department of Physics \& Astronomy, Evanston, IL  60208-3112, USA
}%

\author{W. H. Lippincott}
\thanks{Formerly at 1.
    %Fermi National Accelerator Laboratory (FNAL), Astrophysics Department, Batavia, IL 60510-5011, USA
}
\affiliation{
 University of California, Santa Barbara, Department of Physics, Santa Barbara, CA 93106-9530, USA
}%

\author{A. Monte}
\thanks{Formerly at 1.
    %Fermi National Accelerator Laboratory (FNAL), Astrophysics Department, Batavia, IL 60510-5011, USA
}
\affiliation{
 University of California, Santa Barbara, Department of Physics, Santa Barbara, CA 93106-9530, USA
}%

\author{J. Phelan}
\thanks{Formerly at 2.
    %Northwestern University, Department of Physics \& Astronomy, Evanston, IL  60208-3112, USA
}
\affiliation{%
 Massachusetts Institute of Technology, Department of Physics, Cambridge, MA 02139-4307, USA
}%

\date{September 23, 2021}% It is always \today, today,

\begin{abstract}
%very abstract
Dark matter searches using dual-phase xenon time-projection chambers (LXe-TPCs) rely on their ability to reject background electron recoils (ERs) while searching for signal-like nuclear recoils (NRs). % based on the ratio of ionization to scintillation produced in the recoil event. 
ER response %in dual-phase xenon time-projection chamber (LXe-TPC) dark matter searches
is typically calibrated using $\beta$-decay sources, such as tritium, but
these calibrations do not characterize events accompanied by an atomic vacancy, as in solar neutrino scatters off inner-shell electrons.  Such events lead to emission of x rays and Auger electrons, resulting in higher electron-ion recombination and thus a more NR-like response than inferred from $\beta$-decay calibration. 
%
% Dark matter searches using dual-phase xenon time-projection chambers (LXe-TPCs) rely on their ability to reject background electron recoils (ERs) while searching for signal-like nuclear recoils (NRs) based on the ratio of ionization to scintillation produced in the recoil event. 
% %
% Detector response to low-energy ER events is often calibrated using $\beta$-decay sources, such as tritium, and applied to other sources of ER events, such as neutrino-electron scattering from solar neutrinos.
% %
% In such scattering events, however, the neutrino has appreciable probability of interacting with an inner-shell electron, leaving an atomic vacancy that results in secondary emission of x-rays and Auger electrons.
% %
% This emission of multiple, low-energy secondary particles increases the ionization density, and therefore recombination, at the interaction site, leading to a detector response more akin to nuclear recoils (signal) than a $\beta$-decay calibration would indicate.
% %
% In this work, 
We present a cross-calibration of tritium $\beta$-decays and $^{127}$Xe electron-capture decays (which produce inner-shell vacancies) in a small-scale LXe-TPC and give the most precise measurements to date of light and charge yields for the $^{127}$Xe $L$-shell electron-capture in liquid xenon.
We observe a 6.9$\sigma$ (9.2$\sigma$) discrepancy in the L-shell capture response relative to tritium $\beta$ decays, measured at a drift field of 363 $\pm$ 14 V/cm (258 $\pm$ 13 V/cm), when compared to simulations tuned to reproduce the correct $\beta$-decay response.
%
%The discrepancy is in the direction that supports our hypothesis of increased recombination leading to more NR-like signals.
%
%We also present a model to extend the measurement of L-shell vacancies to L-shell recoils including the primarily ejected electron, which can be included in solar neutrino background models for large-scale LXe-TPC dark matter searches.
%
%This model indicates a 5\% increase in neutrino ER background leakage due to the inner-shell effect.  This has a negligible impact on the sensitivity of LXe-TPC dark matter searches when the background is correctly modeled.  
%
In dark matter searches, use of a background model that neglects this effect leads to overcoverage (higher limits) for background-only multi-kiloton-year exposures, but at a level much less than the 1-$\sigma$ experiment-to-experiment variation of the 90\% C.L. upper limit on the interaction rate of a 50~GeV/$c^2$ dark matter particle. 

%
%\begin{description}
%\end{description}
\end{abstract}

\maketitle

\section{\label{s:intro} Introduction}

\par For over a decade, dual phase xenon time projection chambers (LXe-TPCs) have led the search for particle dark matter in the $\sim$10~GeV--10~TeV mass range~\cite{ref:LUX_DM,ref:PandaX2_DM,ref:XENON1T_DM_SI,ref:XENON1T_DM_SD}, which includes weakly interacting massive particles~\cite{ref:Jungman1996,ref:Roszkowski2018}. The success of the LXe-TPC technique stems from both its scalability and its ability to discriminate against events arising from ambient radioactivity using a combination of self-shielding, position reconstruction, and electron recoil (ER) vs nuclear recoil (NR) discrimination. The ER/NR discrimination is critical to eliminate low-energy, single-scatter ER backgrounds distributed uniformly in the detector, e.g., from the $\beta$ decays of $^{85}$Kr and $^{214}$Pb. The three multiton LXe-TPC experiments that are now underway around the world~\cite{ref:LZ_WIMP,ref:PandaX4T_WIMP,ref:XENONnT_WIMP} are sensitive to an additional low-energy ER background that must be addressed by ER/NR discrimination:  neutrino-electron scattering by solar neutrinos~\cite{ref:Chen2016, ref:LZ_WIMP}. The tacit (and not unreasonable) assumption made by these experiments to date is that the ERs produced by neutrino scatters will have the same, well-calibrated signature as $\beta$ decays. The measurement described in this paper shows that this assumption of a universal ER response does not hold in LXe-TPCs, and a new model for the neutrino-electron scattering response is presented.

\subsection{\label{ss:problem}ER / NR Discrimination and Auger Cascades}

\par Energy depositions in an LXe-TPC generate both prompt scintillation light (S1) and ionization. The ionization drifts under an applied electric field to a gas region where it produces a delayed scintillation signal (S2) via electroluminescence~\cite{ref:Chepel_2013}. The ratio of ionization to scintillation (S2/S1) is, on average, larger for ERs than for NRs, for the same amount of S1 light. That ratio is determined by both the relative amounts of ionization and atomic excitation initially present in the ER (or NR) track and by the fraction of ionization electrons that recombine with xenon ions at the interaction site, simultaneously reducing S2 and increasing S1~\cite{ref:LUX_NRrejection}. 

\par
\begin{figure}[t]
    \centering
    \includegraphics[width=8.6cm]{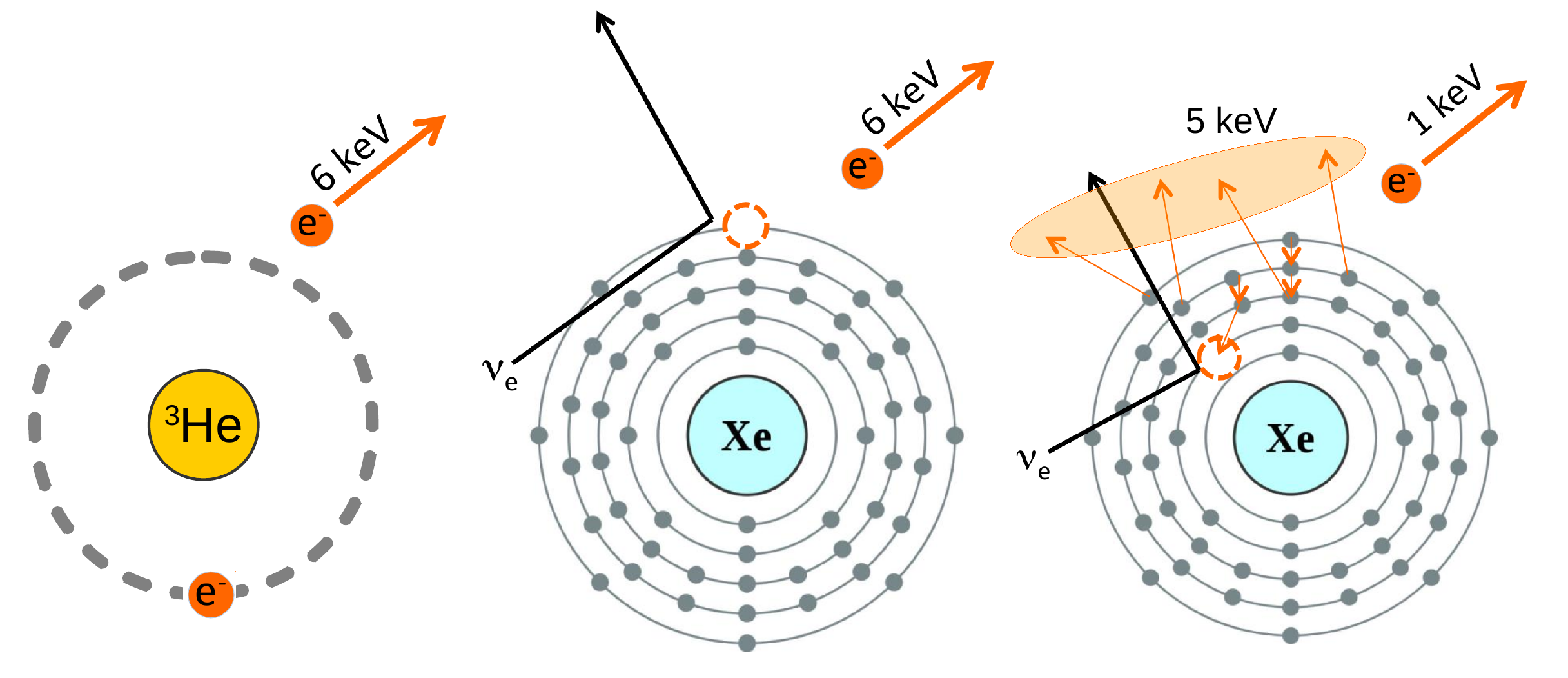}
    \caption{A diagram depicting a $\beta$ decay from tritium (left), a valence recoil from neutrino or Compton scattering (center), and an inner-shell recoil (right). Each of these interactions deposits a total of 6 keV via electron recoil(s), but the inner-shell recoil produces a distinct event topology from either the $\beta$-decay or valence recoil.}
    \label{fig:topology}
\end{figure}

 A neutrino scattering off a valence atomic electron will likely have the same signature as an equal-energy naked $\beta$ decay (i.e., a $\beta$ decay directly to the ground state of the daughter nucleus) because, in both cases, the energy deposited in the detector is given entirely to a single recoiling electron. The story changes when a neutrino scatters off an inner-shell electron, leaving a vacancy in the atomic structure of the struck xenon atom. In this case, the unstable atomic state relaxes by the emission of either an x ray or an Auger electron, with the potential to create additional atomic vacancies. The total energy deposited by these relaxation processes will equal the binding energy of the ionized inner-shell electron. %
Scatters off $L$-shell electrons are of particular interest to LXe-TPC dark matter searches as the xenon $L$-shell energy (4.8--5.4 keV) lies in the middle of the typical 1.5--15~keV$_{\rm ee}$ (electron-equivalent) dark matter region of interest. In xenon, vacancies at the $L$-shell and beyond most frequently relax via Auger emission of an electron from the next shell out, creating a cascade of Auger emission~\cite{ref:auger_cascade,ref:nds_livechart,ref:Ringbom2012}. In this manner, a 6 keV energy deposition by a neutrino scatter on the $L$ shell can be shared between up to eight electrons, all coming from the same xenon atom and all with energies below 2.5~keV. Figure~\ref{fig:topology} shows a schematic of the topology for a tritium $\beta$ decay, a valence-shell scatter in xenon, and an inner-shell scatter. 

\par Because an electron's stopping power increases as the electron energy falls, the Auger cascade event topology leads to a more compact energy deposition than a $\beta$ decay of the same total energy. In technologies such as the bubble chamber, where ER/NR discrimination relies on differences in energy density, this has an orders-of-magnitude impact on discrimination~\cite{ref:PICO_ER}, but how this may impact discrimination in an LXe-TPC is less obvious. The dominant effect driving ER/NR discrimination in LXe-TPCs is the primordial ionization to excitation ratio~\cite{ref:NEST,ref:Dahl_thesis}, but the increased ionization density in an Auger cascade could lead to increased recombination, reducing S2/S1 and causing the event to appear more NR-like. In this work, we show that there is a measurable decrease in the ionization-to-scintillation signature of events with inner-shell vacancies. 

\subsection{\label{ss:concept} Measurement Concept:  $^{127}$Xe electron-capture (EC) decay}

\par The effect of an inner-shell vacancy on recombination (and on the resulting S2/S1 distribution and discrimination power) can be measured by observing $^{127}$Xe electron-capture decays in the LXe-TPC. In this decay process, an s-orbital electron is captured by the nucleus and an electron neutrino is emitted. If the daughter nucleus is in the ground state, the only visible energy in this decay comes from the resulting atomic vacancy. In the case of $^{127}$Xe, the daughter $^{127}$I nucleus is left in an excited 375~keV (203~keV) state with 47\% (53\%) branching ratio. 

\par In a measurement of $^{127}$Xe decays by the LUX experiment, the gamma rays emitted by the excited $^{127}$I nucleus were contained in the detector, creating multisite interactions where the S2, but not S1, generated by the relaxation of the atomic vacancy could be isolated. LUX found the charge yield (S2) for $L$-shell captures to be lower than expected based on calibrations with tritium $\beta$ decays~\cite{ref:LUXXe127}, but low rate and the lack of S1 limited the significance of the measurement. The objective of this work is to observe $^{127}$Xe decays in a small LXe-TPC where the $^{127}$I gammas can escape, collecting events where the sole energy deposition comes from the relaxation of the atomic vacancy (see Table~\ref{tab:xe127}). The resulting, high-statistics measurements of the S1 and S2 generated by $L$-shell vacancies can then be directly compared to tritium $\beta$ decays measured in the same detector. 

\begin{table*}
    \centering
    \begin{tabular}{c|ccc|c}
    \hline
    \hline
         \rule{0pt}{2.5ex}Gamma cascade & \multicolumn{3}{c|}{Intensity by shell [\%]} & $P_\mathrm{esc}$ \\
         $[$keV$]$ & K (33.2 keV) & L (5.2 keV) & M (1.1 keV) & [\% @ 2.0 cm] \\
         \hline
         \hline
%         618.4 &  0.04 & 0.07 & 0.02 & 62.2 \\
%         \hline
%         375.0 & 15.85 & 2.56 & 0.56 & 45.0 \\
%         172.1 + 202.9 & 22.26 & 3.59 & 0.79 & 0.5 \\
%         172.1 + 145.3 + 57.6 & 1.40 & 0.23 & 0.05 & 0.0 \\
%         \hline
%         202.9 & 41.99 & 6.03 & 1.36 & 10.6 \\
%         145.3 + 57.6 & 2.64 & 0.38 & 0.09 & 0.0 \\
         \rule{0pt}{2.5ex}618.4 &  0.0 & 0.1 & 0.0 & 62.2 \\
         \hline
         \rule{0pt}{2.5ex}375.0 & 15.9 & {\bf 2.6} & 0.6 & 45.0 \\
         172.1 + 202.9 & 22.3 & 3.6 & 0.8 & 0.5 \\
         172.1 + 145.3 + 57.6 & 1.4 & 0.2 & 0.0 & 0.0 \\
         \hline
         \rule{0pt}{2.5ex}202.9 & 42.0 & {\bf 6.0} & 1.4 & 10.6 \\
         145.3 + 57.6 & 2.6 & 0.4 & 0.1 & 0.0 \\
         \hline
         \hline
    \end{tabular}
    \caption{$^{127}$Xe decay scheme and gamma escape probabilities. Values in the middle columns give the probability that the decay captures on a given shell (denoted by the columns) and follows the given gamma cascade (denoted by row). The final column gives the probability that the only energy deposited within 2.0~cm of the interaction site comes from the atomic vacancy, i.e., that no internal conversion electrons are emitted and all emitted gammas travel at least 2.0~cm from the decay site without scattering. Roughly 2\% of all $^{127}$Xe decays result in an $L$-shell vacancy with no other visible energy deposition, primarily from the two bold entries in the table. Data taken from Refs. \cite{ref:XCOM,ref:nds_livechart,ref:nist_chem,ref:Rodrigues2014}.}
    \label{tab:xe127}
\end{table*}

\section{\label{s:hw} Experimental Setup}

\subsection{\label{ss:detector} XELDA detector}

\par
\begin{figure}[t]
    \centering
    \includegraphics[width=8.6cm]{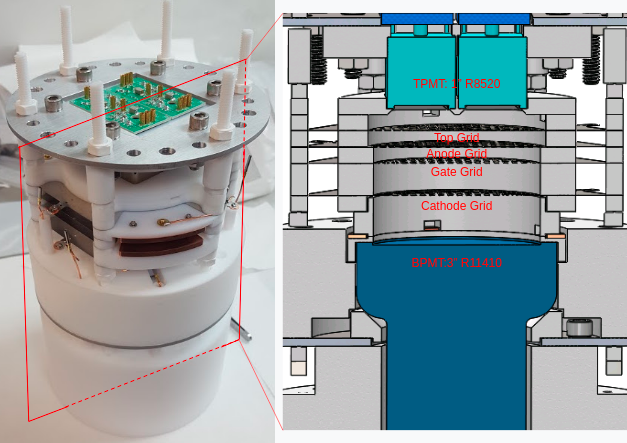}
    \caption{Image (left) and solid model cross-section (right) of the XELDA TPC. In normal operation, the liquid surface sits midway between the anode and gate grids.  The liquid level is measured at three locations via the capacitance between pairs of copper plates -- the bottom plate of one pair is visible in the center of the image on the left.}
    \label{fig:xelda}
\end{figure}

 The Xenon Electron-Recoil $L$-shell Discrimination Analyzer (XELDA) detector is a 6.33-cm-diameter LXe-TPC with a 1.27 cm tall active region, giving a 40 mL (177 g) LXe target (see Fig.~\ref{fig:xelda}). The XELDA design is based on the MiX detector at the University of Michigan~\cite{ref:MiX}. Scintillation in the chamber is detected by five photomultiplier tubes (PMTs):  a single, 3-in.-diameter Hamamatsu R11410 at the bottom of the detector to measure the prompt S1 scintillation, and four 1-in.-square Hamamatsu R8520 PMTs at the top to measure S2 electroluminescence. The S2 hit-pattern in the four top PMTs allows millimeter-resolution reconstruction of the $(x,y)$ position of events in the middle of the detector and robust rejection of events occurring outside the central volume.

\par The main TPC structure is fabricated from white polytetrafluoroethylene (PTFE, or Teflon) to increase light collection efficiency. Electric fields in the chamber are established by four electroformed, stainless steel, honeycomb-pattern grids. From bottom to top these grids are the cathode, gate, anode, and top. The LXe-TPC target lies between the cathode and gate grids, and the liquid xenon surface lies between the gate and anode. Three parallel plate capacitors in the gate and anode grid planes measure the position of the liquid surface. Ionization electrons in the target volume drift upward past the gate grid to the liquid surface and into the gas phase, producing electroluminescence as they traverse the gap between the liquid surface and the anode grid. The anode and gate grids are aligned such that each hexagonal hole in the gate grid lies beneath a vertex in the anode grid, minimizing the spread in gas path lengths of drifting electrons. The top grid establishes a reverse-field region in the gas space above the anode, and a copper ring below the cathode collects ionization electrons produced in the reverse-field region between the cathode and bottom PMT.

\par The grids each have 2 mm pitch (distance between opposite sides of each hexagonal cell) and the grids in the liquid (gas) are 50~$\mu$m (120~$\mu$m) thick. The electroforming process gives the grid wires a hexagonal cross-section, with wires roughly as wide as they are thick. The wire cross sections for each of the two grid types were measured under a microscope, and the as-fabricated grids were modeled in COMSOL to find the electric field leakage across each grid. Two electric field configurations are used in the data presented below, giving calculated TPC drift fields of $363\pm14$~V/cm and $258\pm13$~V/cm, with the dominant uncertainty coming from the position of the liquid surface. The gas extraction field is kept at 9.5~kV/cm in both field configurations. The electron drift times for interactions at the cathode (gate) grids are 8.5 (1.0) $\mu$s.

\subsection{\label{ss:fluids} Cryogenics and Circulation System}

\par
\begin{figure}[t]
    \centering
    \includegraphics[trim=0.58in 0.35in 0.60in 0.35in, clip, width=8.6cm]{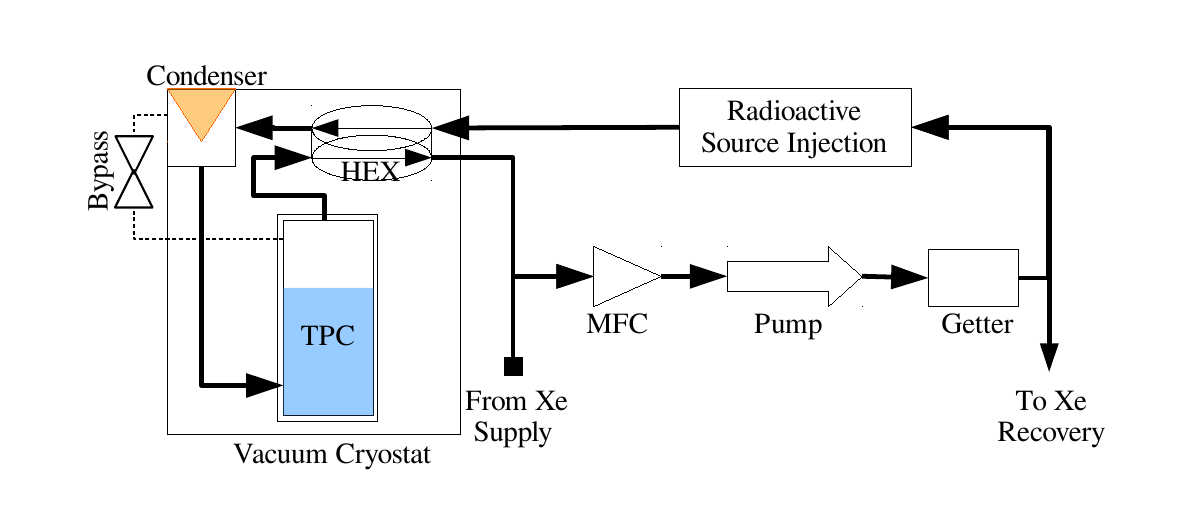}
    \caption{Simplified piping and instrumentation diagram for the XELDA xenon circulation system. The system can operate either in an active circulation mode, where a pump drives xenon through the purification and source injection systems, or in a passive circulation mode, where gas from the TPC is drawn through a bypass to a cold head where it condenses and drips back to the inlet at the bottom of the TPC.}
    \label{fig:PnID}
\end{figure}

 The XELDA TPC hangs inside a 6'' diameter stainless steel vessel, which in turn is housed inside a vacuum cryostat repurposed from the SCENE experiment~\cite{ref:scene}. A xenon circulation system delivers liquid xenon to the bottom of the TPC vessel and removes gaseous xenon from the top, maintaining both the purity of the liquid xenon and the thermodynamic state of the chamber, see Fig.~\ref{fig:PnID}. Both ``active'' and ``passive'' circulation modes are needed to operate the chamber. During active circulation, an all-metal bellows pump (Senior Aerospace MB-111) draws xenon gas from the TPC and drives it through a heated zirconium getter (SAES MonoTorr PS4-MT3-R-1), back to a condenser held at 167.0~K by a cryocooler (Cryomech PT60) and  proportional–integral–derivative-controlled heater. The condensed xenon drips down to the inlet at the bottom of the TPC. Xenon flow in this mode is throttled at 3~slpm by a mass flow controller at the pump inlet, and a heat exchanger between the gas streams leaving and returning to the cryostat keeps the heat load well within the PT60's cooling power. Circulation in this mode can increase electron lifetimes in the TPC from 1 to $>$10~$\mu$s in less than a week of circulation, limited primarily by inefficient mixing of xenon into the TPC.  

\par In active circulation, the liquid level in the TPC fluctuates widely, driven by variations in the differential pressure between the TPC and condenser volumes. To achieve the stable liquid level needed for TPC operation, the pump is turned off and a bypass valve connecting the TPC gas space directly to the condenser is opened. In this mode, the liquid xenon level in both the TPC and condenser drain line is determined entirely by the total xenon mass in the system and the xenon liquid temperature (density). Xenon is not efficiently purified in this mode (the TPC is able to maintain $>$10$~\mu$s electron lifetimes without active purification), but the TPC is still cooled by the incoming xenon as the condenser and TPC effectively form a closed-loop thermosyphon. In both active and passive circulation modes, the TPC temperature is regulated by a proportional–integral–derivative-controlled heater at the bottom of the TPC vessel.

\subsection{\label{ss:sources} Source production and injection}

\par The measurements made with the XELDA detector require radioactive sources to be uniformly distributed within the TPC volume. These sources are injected directly into the gas stream during active circulation, downstream of the heated getter. Tritium was injected in the form of tritiated methane (CH$_3$T), %passing through a XXXXX purifier, 
and was later removed from the system by resuming circulation through the heated getter. The $^{127}$Xe sources were created at Fermilab by activating 25 and 50 mL high-pressure bottles of xenon in Fermilab's Neutron Irradiation Facility~\cite{ref:fnal_ntf,ref:amols_1977}. This activation also produces $^{129m}$Xe (8.9 day half-life), $^{131m}$Xe (11.8 day half-life), and $^{133}$Xe (5.2 day half-life), but $^{127}$Xe has the longest half-life of the activation products at 36.3 days, and after a $\sim$50-day-long cooldown period is the dominant radioactive isotope in the sample. We cannot remove the $^{127}$Xe once injected, so this month-long half-life also dictates much of the run plan described below.

\subsection{\label{ss:electronics} Electronics and Trigger Scheme}

\par
\begin{figure}[t]
    \centering
    \includegraphics[trim=0.55in 0.39in 0.48in 0.39in, clip, width=8.6cm]{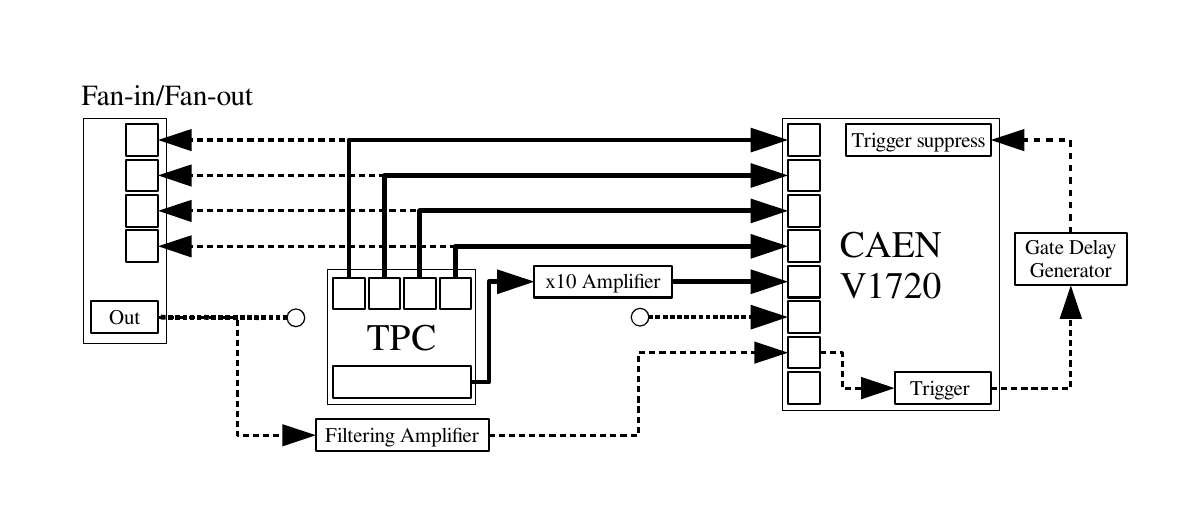}
    \caption{Schematic showing the XELDA digitization and trigger electronics. Signal paths used in analysis are shown by solid lines, trigger paths are shown by dashed lines, and signals used for diagnostics only are shown by dotted lines.}
    \label{fig:electronics}
\end{figure}

 The XELDA data acquisition (DAQ) system deploys a CAEN V1720, eight-channel, 12 bit, 250 MHz digitizer. The bottom PMT signal passes through a $\times$10 amplifier immediately outside the vacuum jacket, and the amplified signal is digitized, ensuring a clear single-photoelectron signal and easy photon counting in small S1 pulses. This amplified signal saturates the digitizer for most S2 pulses in the region of interest, so S2 measurements are made using only the top four PMTs. Signals from the top PMTs are cloned by a nuclear instrumentation module fan-in/fan-out module, with one copy of each PMT signal digitized directly and a second copy going to trigger electronics. A schematic of the data acquisition electronics is shown in Fig.~\ref{fig:electronics}.

\par The XELDA DAQ operates primarily on an S2 trigger, with waveforms recorded for 15~$\mu$s before and after the trigger signal. This 30 $\mu$s window defines a single ``event'' in the TPC. To generate the trigger, a second fan-in/fan-out creates an analog sum of the top four PMTs, and that sum is fed to an Ortec 579 amplifier with 0.5 $\mu$s integration and differentiation times. The CAEN digitizer self-triggers when the amplified sum of the top PMT waveforms exceeds a fixed threshold for 300~ns. This trigger is highly efficient for S2 pulses down to five extracted electrons and has nonzero efficiency for triggering on single-electron pulses, but it is blind to S1s from events up to and beyond the $K$-shell energy. This trigger configuration allows the digitizer to ignore S1-only events from interactions in the reverse-field region below the cathode and in the xenon spaces outside the TPC field cage. Finally, a gate-delay generator enforces a 250 $\mu$s dead time after each trigger, preventing triggers on after-pulsing following large S2 pulses.

\section{\label{s:data}Data Collection, Reduction, and Cross-calibration}

\subsection{\label{ss:datasummary} Summary of data collected}

\par
\begin{figure}[t]
    \centering
    \includegraphics[width=8.6cm]{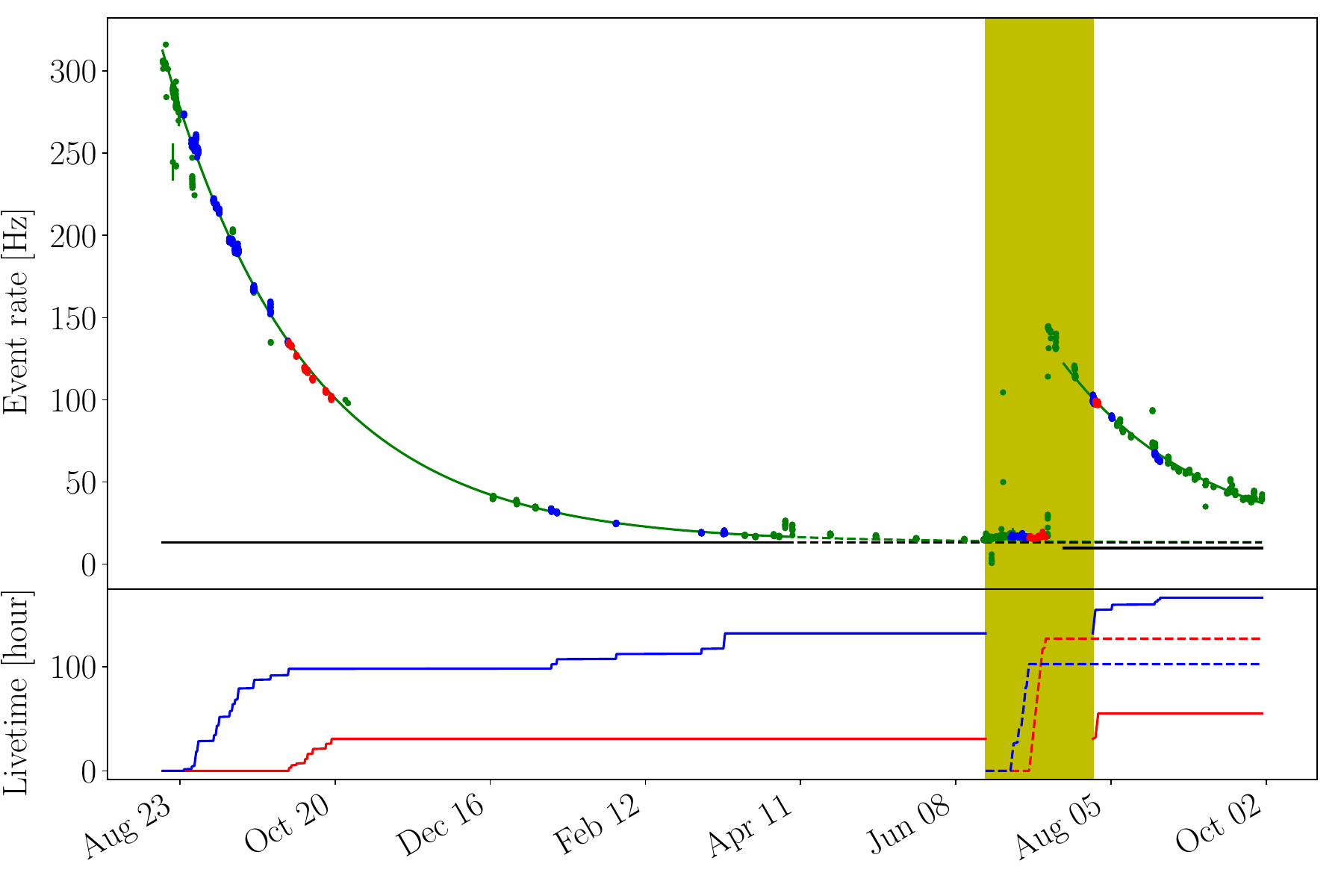}
    \caption{Trigger rates (above) and accumulated live times (below) for data taken by XELDA from August 2018 through October 2019. Above:  blue and red points in the rate plot indicate data used in this paper at 363 and 258~V/cm, respectively. Green points indicate datasets not used in analysis but shown here to illustrate the decay of $^{127}$Xe in the detector. These runs were rejected due to effects including poor electron lifetime ($<$10~$\mu$s), unstable liquid level, lack of clear single- and double-electron peaks in the S2 spectrum, or abnormal trigger rate. Lines indicate fits to the data rate taking a flat background plus decaying exponential with fixed 36.3-day half-life. The shaded band indicates the period where tritium is present in the detector. In the bottom panel, solid (dashed) lines indicate xenon-only (tritium) live times.}
    \label{fig:rates}
\end{figure}

 The XELDA detector acquired data over a period from August 2018 to October 2019. Data were taken with two electric field configurations (drift fields of 363 and 258~V/cm) and two source configurations: ``xenon-only'' data, with trigger rates from $^{127}$Xe of 10--300~Hz, and ``tritium'' data, which include some residual $^{127}$Xe activity as well as a $\sim$2 Hz trigger rate from tritium decay. Figure~\ref{fig:rates} shows the rates, calendar durations, and accumulated live times for each of the four data categories. There are three source injections relevant to this analysis: an initial $^{127}$Xe injection, a tritium injection once the $^{127}$Xe rate had decayed, and a final $^{127}$Xe injection after tritium data taking was complete. 

\par Data were collected using a customized version of the \textsc{daqman} software \cite{ref:daqman}, which also provided the framework for data reduction described in the next section. We do not process events where the signal from any top PMT exceeded the digitization range, removing all $^{127}$Xe events where the associated gammas deposited significant energy in the TPC forward-field regions, as well as $^{127}$Xe $K$-shell capture events near the walls of the detector where the S2 signal falls predominantly in a single channel. $^{127}$Xe $K$-shell captures in the fiducial region of the detector (defined below) and all outer shell $^{127}$Xe capture events are unaffected by this data preselection.

\subsection{\label{ss:pulse} Pulse identification and classification}

\par Offline waveform processing breaks each event into a sequence of S1-like and S2-like scintillation pulses. This is accomplished by scaling each single-PMT waveform to units of vacuum ultraviolet-photons-detected (phd) per time, suppressing baseline noise in each waveform, summing the five waveforms, identifying pulses within the summed waveform, calculating a set of reduced quantities (RQs) for each pulse, and finally using these RQs to classify each pulse as ``S1'', ``S2'', or ``other''. The methods used here were drawn in part from past work on similar scale TPCs including Refs.~\cite{ref:scene,ref:xenon10}. This section gives key technical details for each of these steps.

\par The single-photo-electron (phe) response of each PMT is measured at the start of each data-taking session using a blue light-emitting-diode (LED) coupled by a fiber optic to the TPC, tracking both excursions and general trends in phe size (or gain) in each PMT. Unlike the LED's 470 nm photons, 175 nm xenon scintillation photons frequently liberate two photoelectrons, so a fixed-value scale factor is applied to the LED data to give the response per vacuum ultraviolet phd. This correction is based on measurements by the LUX-ZEPLIN (LZ) Collaboration of typical double-photoelectron emission rates in the R8520 and R11410 PMT models used in XELDA~\cite{ref:LZ_PMT_DPE}, which we take to be 20.5\% and 22.5\%, respectively.

\par The first 3~$\mu$s of each PMTs waveform is pulse free for reconstructable events, and sets the initial baseline for that waveform. The baseline for the remainder of the waveform is built from a rolling 160 ns average of the waveform, until encountering a pulse that differs from the rolling baseline by seven times the rms found in the initial 3-$\mu$s window. The rolling baseline calculation resumes when the waveform again falls within 1$\times$ the rms of the prepulse baseline, and the baseline during the excursion is taken to be the linear interpolation between the baselines found on either side of the pulse. Once the baseline is found for the entire 30~$\mu$s waveform, the baseline is subtracted and all points not within 120~ns of an excursion beyond a fixed threshold are suppressed. The chosen suppression threshold keeps $>$92\% of single-photoelectron pulses.

\par The scaled, baseline-suppressed signals from all five PMTs are summed, and pulses are found in the summed waveform using a combination of 300 ns and 1 $\mu$s top-hat filters. Time windows containing S2-like pulses are found first by subtracting from the 1-$\mu$s-filtered signal the sliding-window-maximum of the 300-ns-filtered signal, i.e., looking for 1 $\mu$s windows that contain significant pulses even after the largest 300 ns pulse within the window is excluded. After the edges of such S2-like pulses are found, S1-like pulses are identified in the remaining regions of the waveform using the 300-ns-filtered signal only.

\par Reduced quantities are calculated for each pulse found, including the pulse area (phd) by channel; the start time of the pulse (time to reach 5\% of total pulse area in the summed waveform); the prompt fractions of the pulse (\texttt{pfXXX} refers to the fraction of pulse area, in the summed channel, reached within \texttt{XXX}~ns of the pulse start time); and the width of the pulse (\texttt{fwYYZZ} refers to the time it takes the summed waveform to grow from \texttt{YY}\% to \texttt{ZZ}\% of the total pulse area). Based on comprehensive hand scanning, a pulse is classified as S1 if \texttt{pf200}$>$83\% and \texttt{fw1050}$<$150~ns, and classified as S2 if \texttt{pf200}$<$83\% and \texttt{fw1050} is between 60~ns and 1.4~$\mu$s. All other pulses are classified as other.

\subsection{\label{ss:cuts} Event reconstruction}

\par Event reconstruction in XELDA is optimized for the accurate reconstruction of single-scatter events in the fiducial volume of the TPC (defined below), and for the removal of events that either are not single-scatters or fall outside the fiducial volume. To start this process, the largest S2 pulse (determined from the total phd in the four top PMTs) is designated as the ``main'' S2, and the largest S1 pulse (determined from phd in the bottom PMT) prior to that S2 is designated as the main S1. Events with no S2 pulses or with no S1 pulses prior to the main S2 are discarded.

\par
\begin{figure}[t]
    \centering
    \includegraphics[width=8.6cm]{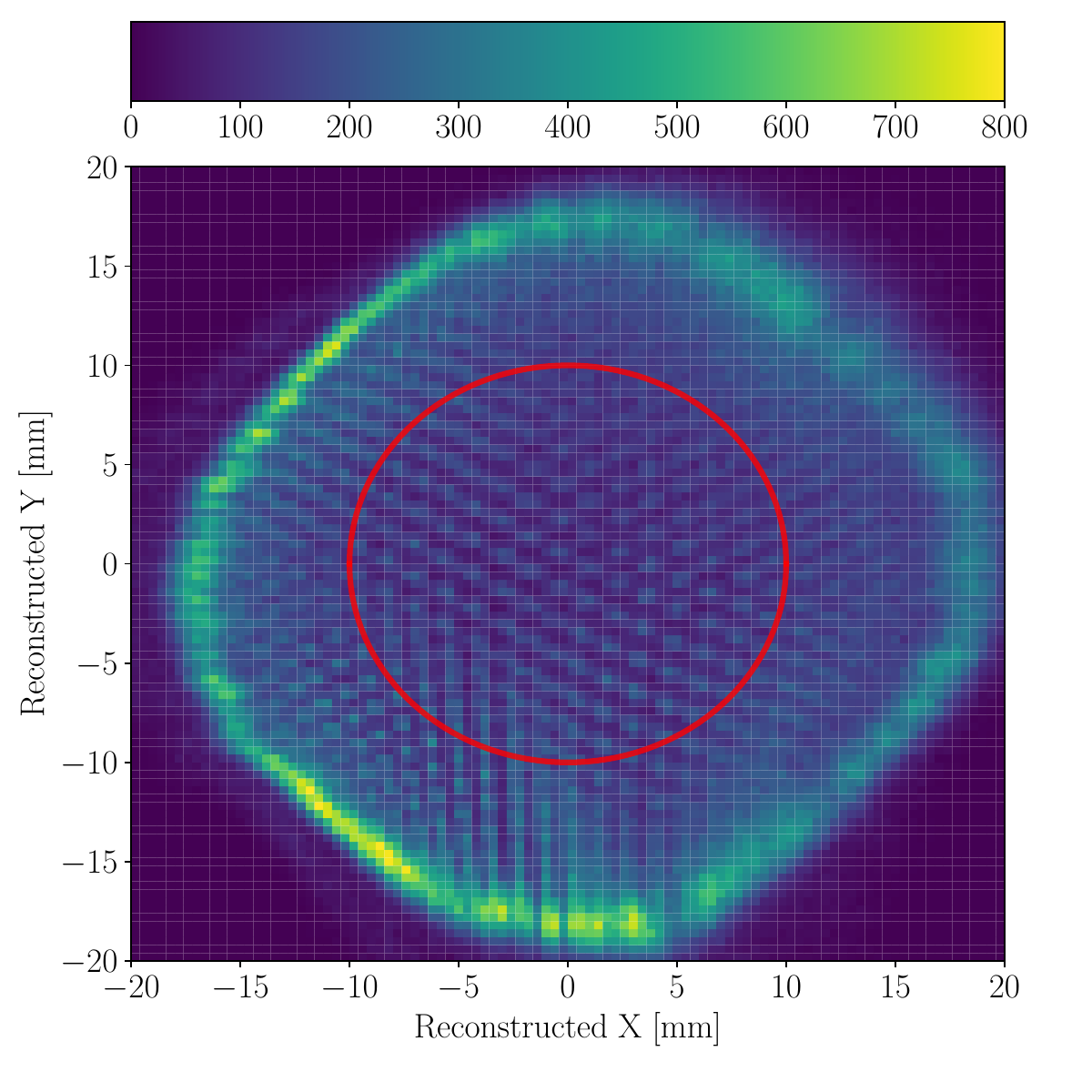}
    \caption{Reconstruction of $(x,y)$ position based on the S2 hit pattern in the top PMTs for $K$-shell events. The honeycomb pattern imprinted by the focusing of electrons as they drift past the gate grid is clearly visible inside the fiducial volume. The red circle indicates the radial extent of the fiducial region of the TPC (10 mm).}
    \label{fig:xy}
\end{figure}

 With the main S1 and S2 pulses identified, three-dimensional (3D) position reconstruction of the event is possible. The drift time separating S1 and S2 (based on the pulse start times defined in the previous section) gives the $z$ position of the event, while the hit pattern of the S2 in the top four PMTs gives the $(x,y)$ position. The $(x,y)$ position is found through a maximum likelihood optimization given an $(x,y)$-dependent light response function (LRF) for each PMT. The LRFs are parameterized as in Ref.~\cite{ref:ZEPXY}, with LRFs for the four PMTs differing only by 90$^\circ$ rotations and an overall scaling to account for tube-to-tube variations in quantum efficiency (QE). Both LRF parameters and relative QEs are found using $L$-shell capture events, selected based on S2 size and assumed to uniformly populate the TPC volume. Relative QEs are fixed by events at the centroid of the $(x,y)$ distribution, assumed to have equal geometric light collection efficiency for all four top PMTs. LRF parameters are found iteratively, using reconstructed positions to generate a new LRF until the LRF parameters stabilize. Finally, two (adjacent) top PMT tubes show nonlinear response for ionization signals beginning around the $K$-shell capture peak. For the purposes of $(x,y)$ reconstruction only, a quadratic correction determined from events at the centroid of the $K$-shell $(x,y)$ distribution is applied to the measured signal size in the two saturating tubes. The accuracy and resolution of the final $(x,y)$ reconstruction is evident in the reconstructed honeycomb pattern of the gate grid using $K$-shell events, shown in Fig.~\ref{fig:xy}.

\par Two pulse-area corrections are applied during event reconstruction, both affecting S2 pulses only. An exponential electron lifetime correction, based on the observed drift-time dependence of uncorrected $K$-shell S2s, scales all S2 pulses to the zero-drift-time value. S2 signals in the top PMTs are also corrected for the relative QEs found in $(x,y)$ reconstruction, with one tube receiving a unity correction. All references below to S2$_\mathrm{top}$ and S2$_\mathrm{CH12}$ include these corrections. No additional corrections (e.g., position-dependent corrections) were found to improve resolution in either S1$_\mathrm{bottom}$ or S2$_\mathrm{top}$ for events in the fiducial volume. PMT nonlinearity does affect S2$_\mathrm{top}$ for $K$-shell capture events, but this is handled during the cross-calibration described in Sec.~\ref{ss:cal} rather than as an explicit correction.

\par The final selection of ``golden'' single-scatter events is based on the following set of cuts. The fiducial volume is set first by requiring the drift time of the event to fall between 2.5--7.5~$\mu$s, avoiding regions with nonuniform electric field near the cathode and gate grids. Next, the reconstructed radial position of the event must be $<$10~mm from the central axis of the detector, eliminating events near the walls of the detector where drifting charge may be lost and $(x,y)$ reconstruction is challenging. Events must fall within the region of interest in (S1$_\mathrm{bottom}$, S2$_\mathrm{top}$) space, defined as 2.0--500~phd in S1$_\mathrm{bottom}$ and 500--70000~phd in S2$_\mathrm{top}$.  No pulses may fall in the first 3~$\mu$s of the event, and the rms of each PMT waveform in that 3~$\mu$s must be $<$1.5 analog-to-digital converter counts, eliminating events with poorly characterized baselines. Multiple-scatter events in the TPC are eliminated by requiring events to contain no pulse larger than 10\% of the main S2. Events with misidentified S1s (due to a particular sporadic electronic pickup problem) are eliminated by requiring S1$_\mathrm{top}$/S1$_\mathrm{bottom}<$3. For normal events this ratio is $<$0.1. Finally, various anomalous event topologies are eliminated by requiring that no more than 20\% of the total event area be contained in pulses outside the main S1 and S2, no more than 5\% of the total event area be contained in pulses classified as ``other'', and no pulses fall between the main S1 and S2 except for single-electron S2s (identified by pulse size and time profile) and S1s smaller than 10\% of the main S1. 

\subsection{\label{ss:cal} Cross-calibration of datasets}

\par The long calendar duration of the XELDA data-taking campaign makes it essential to cross-calibrate data taken at different times and in different detector conditions. In particular, changes in liquid level due to fluctuations in total xenon payload affect electroluminescence production (S2) and potentially the light collection efficiency in the bottom PMT (S1). These and other phenomena affecting S1 and S2 yields are addressed by using the $^{127}$Xe $K$-shell capture peak as a \textit{common candle} in all XELDA data. At 33.2~keV, this peak sits above the tritium $\beta$-decay end point, but is low enough in energy to avoid saturation effects in most (three of five) of the PMTs. We emphasize that we do not use the $K$-shell events to study any aspect of ER discrimination, but instead use it solely as a common point of reference that is easily visible in both types of data that we collected. 

\par The tritium data used in this analysis was taken in a single continuous period (see Fig.~\ref{fig:rates}), and we see no evidence of variation in the $K$-shell peak position or in the tritium $\beta$ continuum, in either the 363 or 258~V/cm tritium datasets. By contrast, both gradual and discrete changes in $K$-shell S1 and S2 were seen in the xenon-only data, so run-by-run (roughly day-by-day) S1$_\mathrm{bottom}$ and S2$_\mathrm{top}$ correction factors are applied to xenon-only data, aligning each run's $K$-shell peak with that observed in the tritium data at the corresponding drift field.

\par The dominant effect seen in the S1$_\mathrm{bottom}$ amplitude is a gradual reduction in $K$-shell signal size, dropping 10\% by the end of the first three months of data taking. This loss of signal appears to be a reduction in the bottom PMT's QE, consistent with aging studies performed on the same tube model at Brown~\cite{ref:brown_pmt} given the $>$100~C of charge collected at the PMT anode during the initial high-rate period. This effect plateaus in later data, as the event rate falls and the PMT aging process slows. Correction factors for this effect are derived from a smoothing spline applied piecewise in calendar time to run-by-run Gaussian fit means of the $K$-shell S1 peak. The smoothing spline avoids over-fitting to statistical fluctuations in the peak position, with breaks between splines inserted by hand to accommodate extended periods when the PMT was unbiased, slowing the aging process. The average absolute residual between the spline and individual S1$_\mathrm{bottom}$ fit means in the $^{127}$Xe data is 0.25\%.

\par Correcting for variation in the S2$_\mathrm{top}$ signal is more complicated because of the nonlinear response at $K$-shell energies in two of the top PMTs. Fortunately, we see no evidence for differences in S2 variation across individual tubes (consistent with the hypothesis that liquid level fluctuations are the primary cause of S2 drift), so we use the signal in the unsaturated top PMT channels 1 and 2, denoted S2$_\mathrm{CH12}$, to make the cross-calibration. 

\par We first scale all xenon-only data to a representative xenon run at each drift field, then find the correction factor between the combined xenon-only $K$-shell peaks and the tritium $K$-shell peaks. Run-by-run xenon-only S2 corrections are made by selecting events whose S1 falls within the $K$-shell region of interest and then finding the S2 scale factor that best aligns the S2$_\mathrm{CH12}$ distribution with a reference distribution, determined by maximizing the $p$ value returned by a Kolmogorov–Smirnov (KS) test. The reference xenon-only S2$_\mathrm{CH12}$ distribution is given by a high-statistics run taken prior to tritium injection, and the maximum correction required in the pre-tritium xenon-only runs is 8\%. Xenon-only runs after tritium was removed exhibit much smaller S2s, requiring 20\%--33\% corrections, consistent with an expected decrease in S2 yield from the increased liquid level following source injections (see below). 

\par
\begin{figure}[t]
    \centering
    \includegraphics[width=8.6cm]{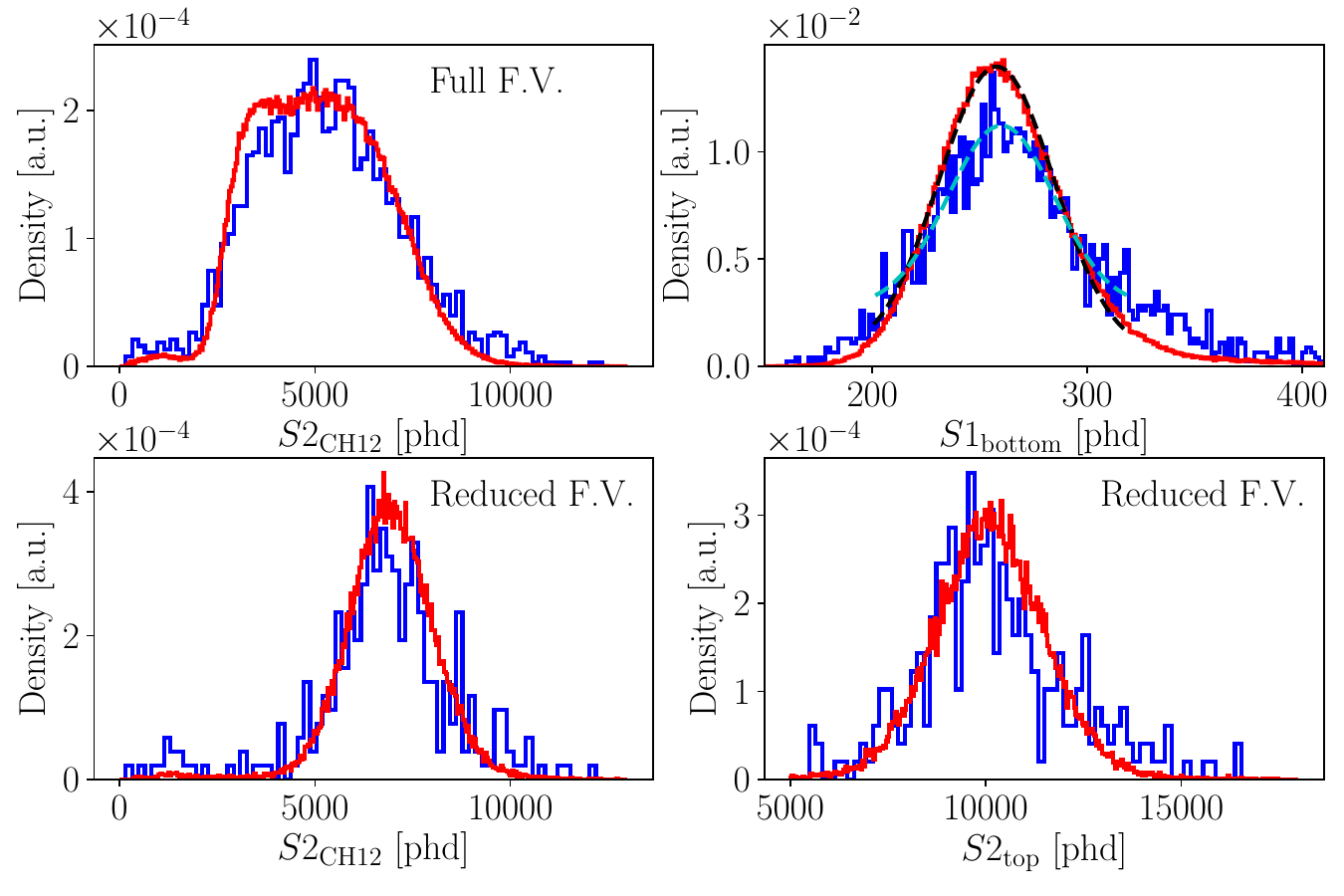}
    \caption{Comparison of the $K$-shell common candles at 363~V/cm, reconstructed after cross-calibration between xenon-only and tritium datasets is complete. In each figure, the red histogram shows the xenon-only distribution and the blue histogram shows the tritium distribution. Dashed curves show Gaussian fits to the S1$_\mathrm{bottom}$ distributions. The top figures use the full fiducial volume, while the bottom two use a reduced volume that is offset from the center of the detector to avoid PMT saturation effects. The vertical axis is scaled such that all the histograms integrate to unity.}
    \label{fig:Kshell}
\end{figure}

 Once the xenon-only runs are made self-consistent, the relative S2 scale between the xenon-only and tritium datasets is found by again maximizing the $p$ value returned by a KS-test. Because S2$_\mathrm{CH12}$ exhibits a strong position dependence (unlike S2$_\mathrm{top}$), the S2$_\mathrm{CH12}$ $K$-shell spectrum in the fiducial volume is rather broad (see Fig.~\ref{fig:Kshell}), and the KS-test is influenced by an unmodeled gamma background that appears beneath the relatively low-rate $K$-shell peak in the tritium data. Therefore, a floating uniform background component is added to the xenon-only S2$_\mathrm{CH12}$ distribution in order to match the observed tritium distribution. Optimized $p$ values of 0.692 (0.623) are found for the 363~V/cm (258~V/cm) datasets, giving a relative scaling between the xenon and tritium data of 0.8235 (0.7945), similar to the drop in S2 seen in going from pretritium xenon-only data to post-tritium xenon-only data.

\par The changes in ionization gain found above are correlated with discrete changes in the  liquid level in the TPC, which result from source injections and xenon recovery events.  In the pretritium xenon data, the liquid height as measured by the capacitive level sensors was 23$\pm$5\% of the distance from the gate to anode.  The liquid surface in the tritium and post-tritium xenon data was at 45$\pm$10\% of the gate-anode gap.  This reduction in the height of the electroluminescence region (with the corresponding increase in the extraction field) gives an estimated $\sim$15\% reduction S2 gain~\cite{ref:NEST}, confirming the expectation that liquid-level shifts are the dominant contributor to the observed changes in S2 gain.

\par Liquid-level shifts also affect the TPC drift field, indicating that the pretritium xenon data had a $\sim$20-V/cm lower drift field (at both field settings) than the corresponding tritium and post-tritium data.  The $K$-shell common candle alignment described above implicitly corrects for this field shift, with the caveat that the $L$-shell peak exhibits slightly less field-dependence than is seen in the $K$-shell.  This is evident in the ratio of the $K$- to $L$-shell S1$_{\mathrm{bottom}}$, which drops from 7.7 at 258~V/cm to 7.5 at 363~V/cm -- that is, the $K$-shell events show more ``field quenching'' in S1 than $L$-shell events.  By relying on the $K$-shell common candle, we over-correct for the effect of any changes in the drift field on the $L$-shell signal, causing us to potentially underestimate the significance of the inner-shell vacancy effect.

\par Whether from variation in ionization/scintillation gain or drift field, uncertainties in the S1 and S2 scale factors between the xenon and tritium datasets dominate the systematic uncertainty in the analysis that follows. The S2 uncertainty is estimated by bootstrapping from the S2$_\mathrm{CH12}$ distribution and building a Neymann construction from the ratio of the global maximum $p$ value described above to the $p$ value obtained using simulation truth information. This construction finds that the true scale factor's $p$ value is greater than 0.6 times the maximum $p$ value in 68\% of trials. Applying this condition to the scale factors tested in our data and marginalizing over the unknown background component gives a relative uncertainty of 1.2\% (0.85\%) for the 363 V/cm (258 V/cm) xenon-to-tritium S2 scale factor.
A similar method is employed for the S1$_\mathrm{bottom}$ channel, giving relative uncertainty of 0.84\% (1.1\%) in the S1 scale factor.

\par Figure~\ref{fig:Kshell} shows the xenon-only and tritium S1$_\mathrm{bottom}$ and S2$_\mathrm{CH12}$ $K$-shell distributions at 363~V/cm after the above scaling is complete. As a cross check, we also compare the full S2$_\mathrm{top}$ distributions for events falling in a reduced fiducial region that selects events where most S2 light goes to the non-saturating PMTs. In all cases a consistent $K$-shell common candle is observed between the cross-calibrated xenon-only and tritium datasets.

\section{\label{s:results}Results}

%\subsection{\label{ss:compare}Direct L-shell to Tritium Comparison}

\par
\begin{figure}[t]
    \centering
    \includegraphics[width=8.6cm]{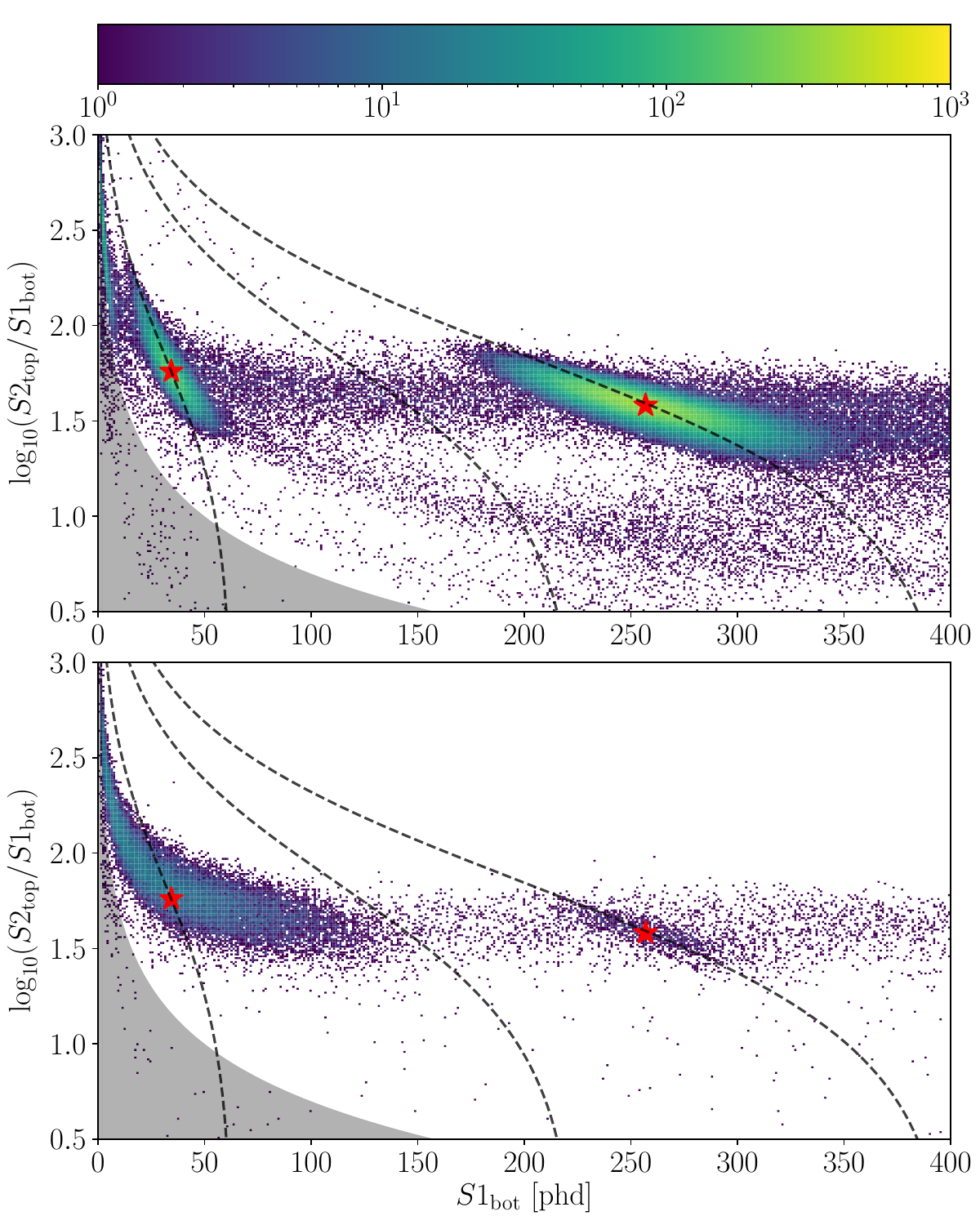}
    \caption{Distribution of $\log_{10}(S2_\mathrm{top}/S1_\mathrm{bot})$ vs. $S1_\mathrm{bot}$ for the 363 V/cm Xe-only data (top) and $^{127}$Xe plus tritium data (bottom). The dashed curves denote constant energies at 33.2 keV ($K$-shell), 18.6 keV (tritium $Q$-value), and 5.2 keV ($L$-shell). The color represents the absolute counts per bin, and the scale is logarithmic. The red stars indicate the centroids of the $K$- and $L$-shell distributions, as determined by the Xe-only data set. Those points are reproduced in the tritium data to help guide the eye. The $L$-shell centroid is clearly offset below the center of the tritium band. The grey regions indicate the analysis threshold, and events in this region are excluded from the analysis.}
    \label{fig:S1S2}
\end{figure}

\begin{figure}[t]
    \centering
    \includegraphics[width=8.6cm]{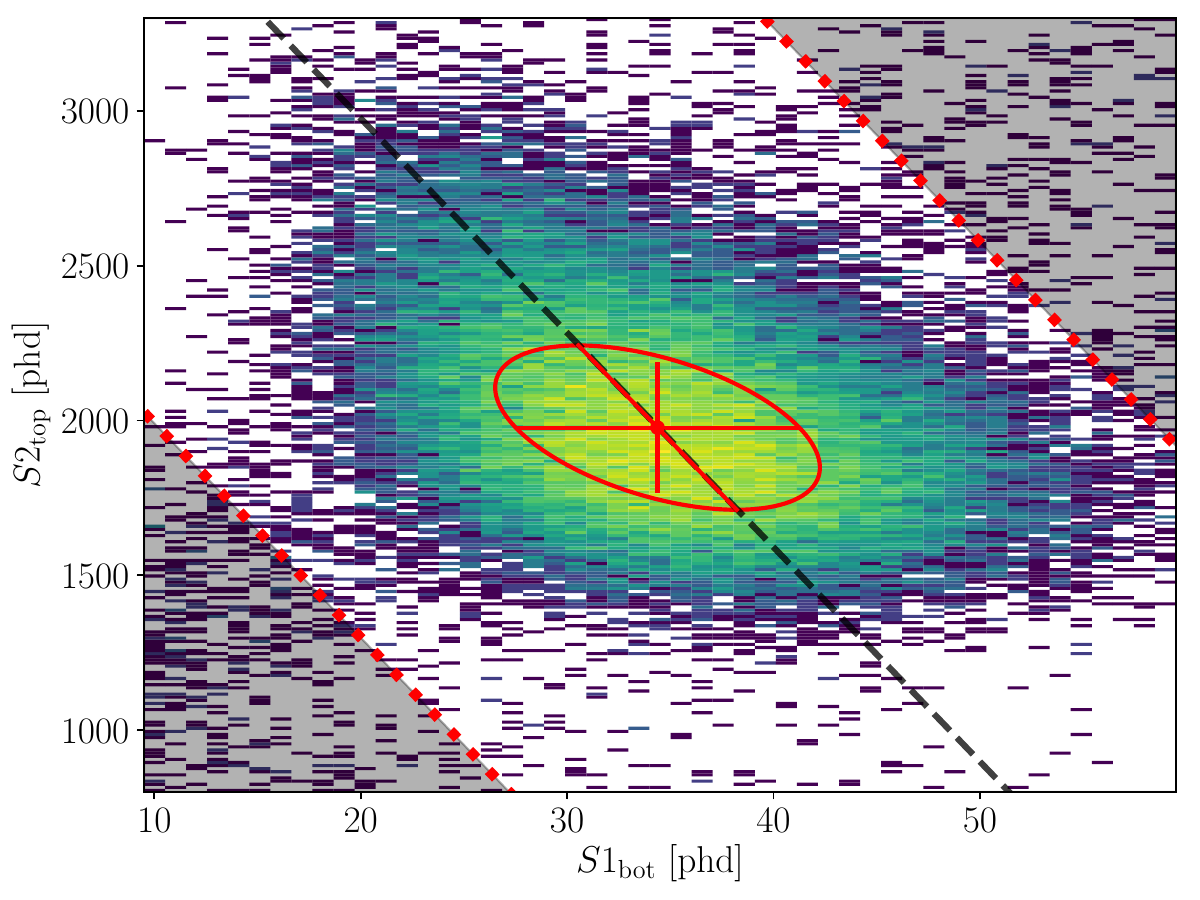}
    \includegraphics[width=8.6cm]{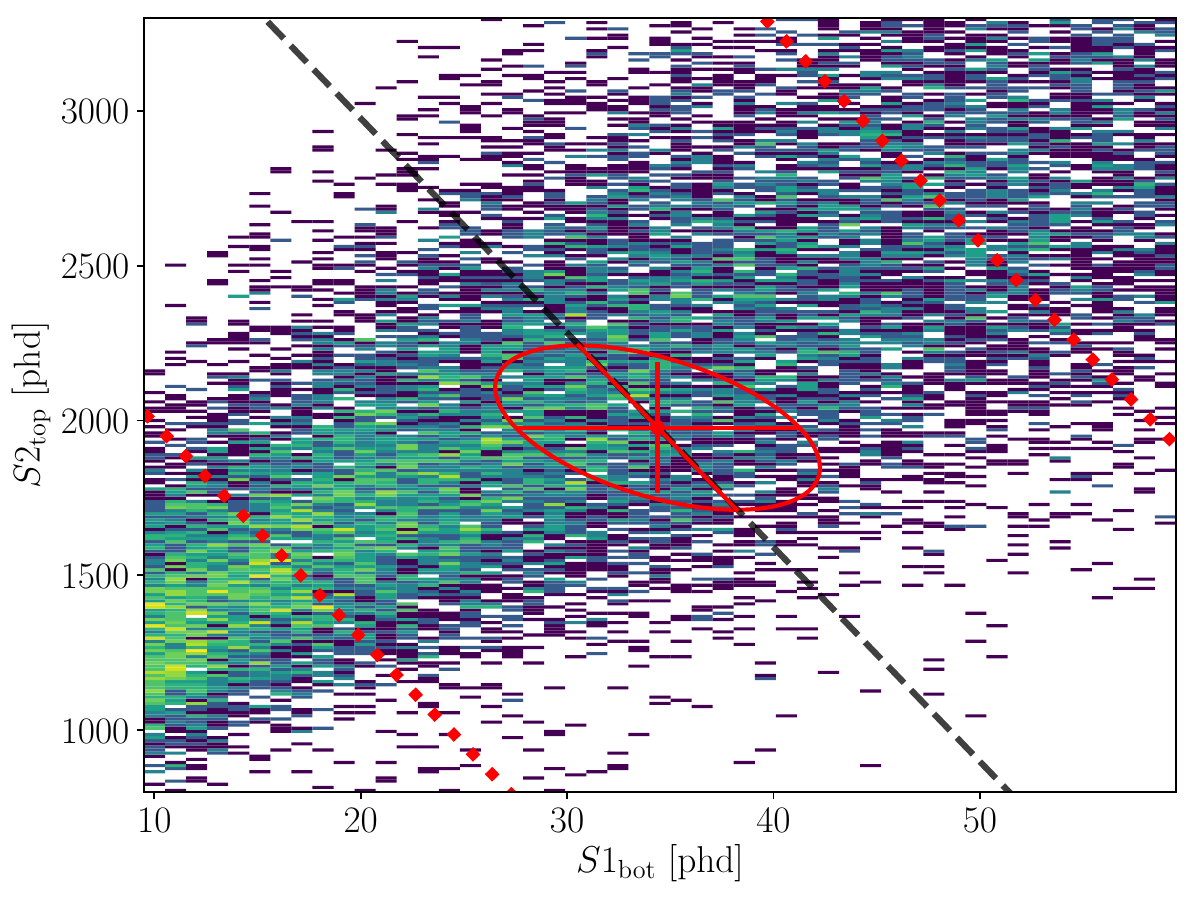}
    \caption{(Top) The $^{127}$Xe $L$-shell electron capture peak, as observed in the 363 V/cm data. The red dotted lines denote the limits of the energy region of interest (the shaded regions are not used in $L$-shell analysis), while the black dashed line denotes the 5.2 keV energy of this peak. The red point is the median in both $S1_\mathrm{bottom}$ and $S2_\mathrm{top}$ for events falling within the energy region of interest (3.20--7.20 keV), while the error bars shown denote the widths ($\sigma$) of the distribution decomposed into independent physical fluctuations: S1 fluctuations (horizontal), S2 fluctuations (vertical), and recombination fluctuations (diagonal). (Bottom) The $^{3}$H $\beta$-decay continuum in the $L$-shell energy region of interest, as observed in the 363 V/cm data, with the same lines of constant energy. The 1$\sigma$ widths of the $L$ shell are drawn on top, again to illustrate the offset.}
    \label{fig:S1S2zoom}
\end{figure}

 Figures~\ref{fig:S1S2} and~\ref{fig:S1S2zoom} show the two-dimensional (2D) [S1, $\log_{10}$(S2/S1)] and (S1, S2) distributions of the 363 V/cm xenon-only and tritium datasets after all cuts and corrections described in the previous section are complete. Visually, it is clear that the $L$-shell peak location is offset from the centroid of the tritium ER band, but it is conceivable that this offset is due to the falling tritium energy spectrum, since the tritium recombination fraction increases (S2/S1 falls) as energy increases. 

\subsection{\label{ss:gains}Simulation and absolute gain calibration}

\par To determine whether the observed offset indicates a true shift in S2/S1 response, we would ideally compare $L$-shell events to an equal-energy $\beta$ decay. Since we do not have a monoenergetic $\beta$ source, we use the Noble Element Simulation Technique (NEST~\cite{ref:NEST} \cite{ref:NEST_Paper}) software package to bridge the gap. We first tune the NEST detector parameters to find agreement between our tritium data and a tritium simulation in NEST, and then use NEST to generate monoenergetic 5.2 keV electron recoils. 

To correctly simulate the statistical variation in signal size, NEST requires the gains of the S1 and S2 signals, defined as
\begin{equation}
\label{eq:gains}
    \textrm{S}1 = g_1 n_\gamma,\quad\textrm{S}2 = g_2 n_e,
\end{equation}
where $n_\gamma$ and $n_e$ are the numbers of scintillation photons and ionization electrons extracted from the interaction site. We use the corrected S1$_\mathrm{bottom}$ and S2$_\mathrm{top}$ as S1 and S2 in Eq.~(\ref{eq:gains}), deriving a single $g_1$ ($g_2$) at each drift field representing the photons detected per photon (electron) emitted in the tritium dataset.

\par The gains $g_1$ and $g_2$ can also be used to construct a linear, drift-field-independent ER energy scale, given by~\cite{ref:Chepel_2013}
\begin{equation}
    \label{eq:keVee}
    E_{ee} = \left(\frac{\textrm{S}1}{g_1}+\frac{\textrm{S}2}{g_2}\right) \times 13.7\textrm{ keV},
\end{equation}
where the subscript on $E_{ee}$ indicates the reconstructed electron-recoil equivalent energy.

\par
\begin{figure}[t]
    \centering
    \includegraphics[width=8.6cm]{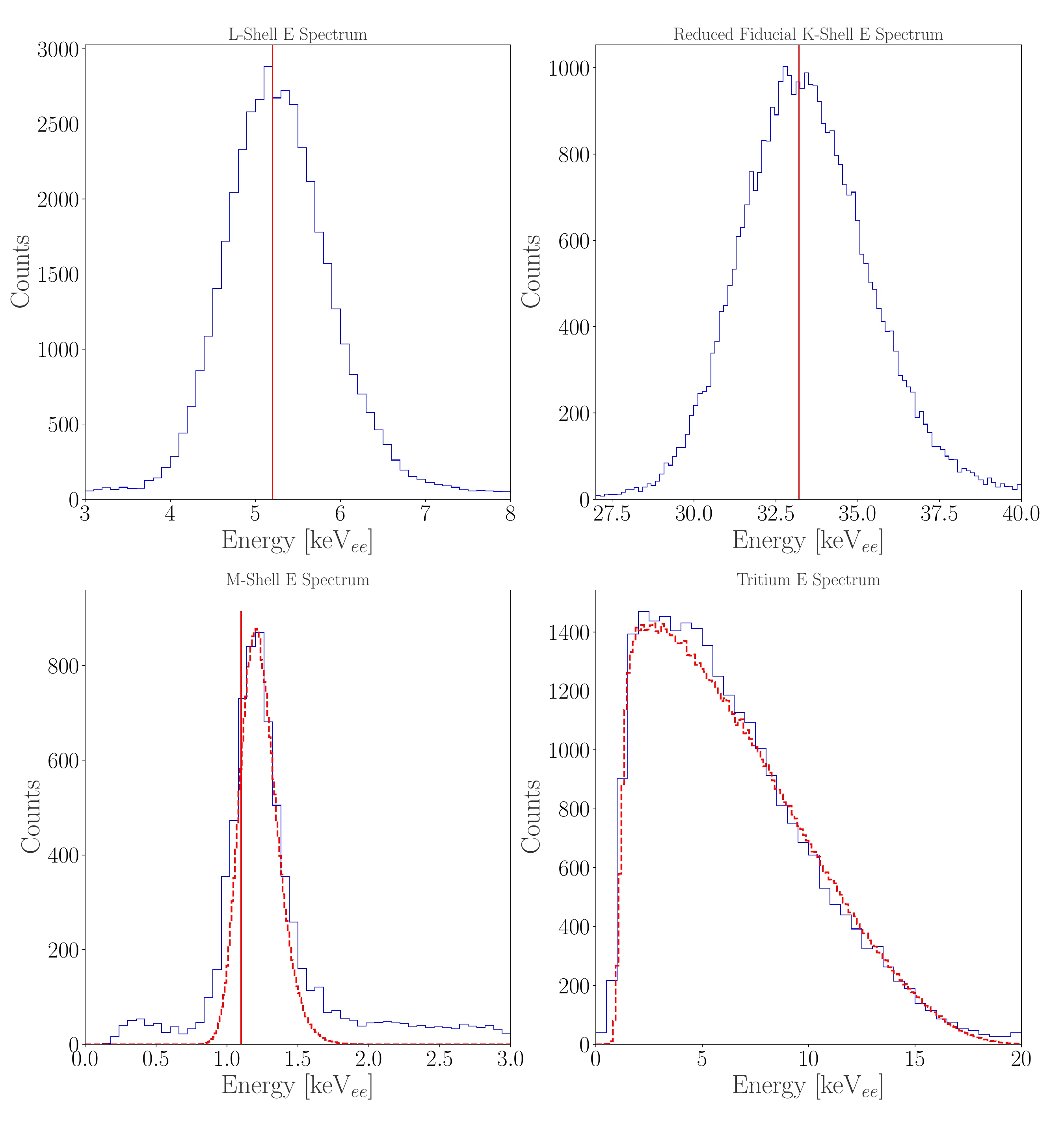}
    \caption{Reconstructed energy distributions for the $^{127}$Xe electron capture peaks and tritium continuum (bottom-right) from 363 V/cm data (blue). Vertical red lines denote the true energy of the $K$- (top-right) and $L$-capture (top-left) peaks (red). The NEST expectation for the tritium energy spectrum (red, bottom-right) in XELDA, including our two-phd S1 threshold, is shown for comparison to the observed data. The NEST expectation for a 1.2 keV $\beta$, also with a two-phd S1 threshold applied, is overlaid on the $M$-shell peak (bottom-left). The $K$-shell peak is shown for a reduced fiducial volume where saturation effects are not present.}
    \label{fig:Espectra}
\end{figure}

\begin{figure*}
    \centering
    \includegraphics[width=12.9cm, trim=2.5cm 0 2.5cm 0, clip]{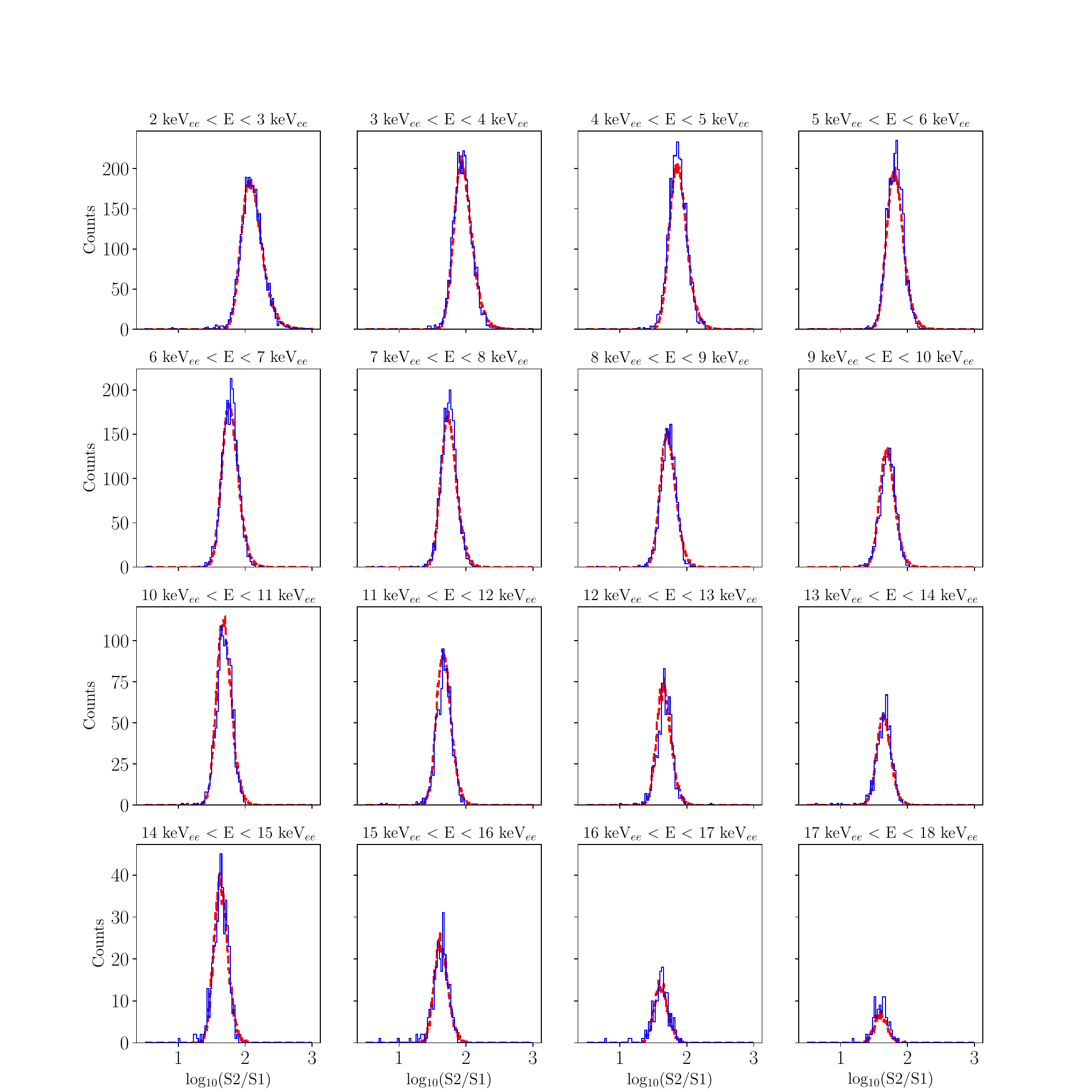}
    \caption{Distributions of $\log$(S2/S1) for the tritium $\beta$-decay continuum, in 1 keV energy bins, comparing the observed tritium data (blue) to expected distributions from a NEST simulation (red) at 363 V/cm. Threshold effects become relevant for signals below 2 keV, so those bins are excluded from this analysis. Above this energy, the observed and simulated distributions are in good agreement. The slight excess in data in the 3-8 keV bins is caused by residual $^{127}$Xe $L$-shell activity in the tritium dataset.}
    \label{fig:tritium_profile}
\end{figure*}

 We find $g_1$ and $g_2$ independently at each field by matching the observed tritium distribution to that simulated by NEST, performing a binned maximum likelihood fit (with Poisson statistics) on the 2D [S1, $\log$(S2/S1)] distribution. The S1 window containing the $L$-shell peak (15--60 phd) is excluded from the fit to avoid contamination by $L$-shell events in the tritium dataset. We also use Eq.~(\ref{eq:keVee}) to constrain $g_1$ and $g_2$ by requiring the $L$-shell centroid to reconstruct to 5.2~keV, leaving a single free parameter for the maximum likelihood fit. The best-fit $g_1$ and $g_2$ with the dominant systematic uncertainties are given in Table~\ref{tab:g1g2}, and the values obtained at each drift field agree to within the uncertainties. Figure~\ref{fig:Espectra} shows the reconstructed energy spectra based on these $g_1$ and $g_2$ values for tritium decays and $K$-, $L$-, and $M$-shell electron capture decays. A comparison of $\log$(S2/S1) between the tritium data and the best-fit simulated tritium profile at 363~V/cm is shown in Fig.~\ref{fig:tritium_profile}. 

\begin{table}[t]
    \centering
    \begin{tabular}{@{\hskip 0.1in}c@{\hskip 0.1in}|@{\hskip 0.1in}c@{\hskip 0.1in}|@{\hskip 0.1in}c@{\hskip 0.1in}}
    \hline
    \hline
    \rule{0pt}{2.5ex}Drift Field & $g_1$ & $g_2$ \\
    $[$V/cm$]$ & $[$phd / photon$]$ & $[$phd / electron$]$ \\
    \hline
    %\hline
    \rule{0pt}{2.5ex}363 & 0.166 $\pm$ 0.001 & 11.5 $\pm$ 0.001 \\
    258 & 0.163 $\pm$ 0.002 & 11.5 $\pm$ 0.002 \\
    \hline
    \hline
    \end{tabular}
    \caption{Observed $g_1$ and $g_2$ values at the two drift fields considered in this analysis.}
    \label{tab:g1g2}
\end{table}

All subsequent references in this work to quantities simulated using NEST use these best-fit XELDA detector parameters. With these values in hand, we can also transform to ``physical" axes, by dividing out the detector-specific gains. In this case, the discrimination parameter becomes $\log(n_e/n_\gamma)$ rather than $\log$(S2/S1).

\par We next find the light and charge yields of $L$-shell electron-capture events and compare to the yields found for monoenergetic $\beta$ decays simulated in NEST. %, using the $g_1$ and $g_2$ derived from the tritium data. 
We define these quantities as
\begin{equation}
    \mathcal{Q}_y = \left\langle\frac{\textrm{S}2}{g_2 E_{ee}}\right\rangle,\quad\mathcal{L}_y=\left\langle\frac{\textrm{S}1}{g_1 E_{ee}}\right\rangle,
\end{equation}
where $\langle\cdots\rangle$ indicates an average taken over a fixed window in reconstructed energy, 3.2~keV~$<E_{ee}<$~7.2~keV. We also define the ratio
\begin{equation}
    q = \quad \frac{\mathcal{Q}^\textrm{L}_y}{\mathcal{Q}^{^3\textrm{H}}_y},
\end{equation}
where the denominator gives the tritium charge yield averaged over the same reconstructed energy window. The advantage of the ratio $q$ is that the $g_1$ and $g_2$ factors enter only via their effect on the energy window and on signal resolution in the NEST simulation, significantly reducing systematic uncertainty.

\par Table~\ref{tab:lshell_yields} shows the charge yields, light yields, and $q$ ratios observed in XELDA data and NEST simulation at both drift fields. Uncertainties on the NEST values are driven by uncertainty on the drift field strength. Systematic uncertainties on the values derived from XELDA data include uncertainties from the cross-calibration of the xenon-only and tritium datasets as well as, for $\mathcal{Q}_y$ and $\mathcal{L}_y$, uncertainties on the gains $g_1$ and $g_2$. The $q$ ratios show a 6.9 (9.2)-$\sigma$ discrepancy at 363 (258)~V/cm between the $L$-shell response and the response expected for a $\beta$ decay.% at the same energy.

\subsection{\label{ss:r-model}Recombination model for $L$-shell capture}

\begin{table*}[t]
    \centering
    \resizebox{\textwidth}{!}{%
    \begin{tabular}{lc|c|c|c}
    \hline
    \hline
    \rule{0pt}{2.5ex}& Source & $\mathcal{Q}_y$ $[e$/keV$_{\rm ee}]$ & $\mathcal{L}_y$ $[\gamma$/keV$_{\rm ee}]$ & $q$ \\
    \hline
    \hline
    \rule{0pt}{2.5ex}XELDA & $^{127}$Xe-L, 363~V/cm  & 33.63 $\pm$ 0.03 (stat) $\pm$ 0.33 (sys)  & 39.36 $\pm$ 0.33 (stat) $\pm$ 0.36 (sys)  & 0.917  $\pm$ 0.001 (stat) $\pm$ 0.009 (sys) \\
    NEST & $\beta$, $r=0.4789$      & 36.42 $\pm$ 0.14                          & 36.44 $\pm$ 0.13                          & 0.9766 $\pm$ 0.0006 \\
    NEST-mod & $\beta$, $r=0.5196$  & 32.98 $\pm$ 0.02                            & 40.02 $\pm$ 0.02 &  0.901 $\pm$ 0.004 \\
    \hline
    \rule{0pt}{2.5ex}XELDA & $^{127}$Xe-L, 258~V/cm  & 32.87 $\pm$ 0.07 (stat) $\pm$ 0.37 (sys)  & 40.12 $\pm$ 0.07 (stat) $\pm$ 0.37 (sys)  & 0.909  $\pm$ 0.003 (stat) $\pm$ 0.007 (sys) \\
    NEST & $\beta$, $r=0.4984$      & 35.10 $\pm$ 0.23                          & 37.90 $\pm$ 0.23                          & 0.9753 $\pm$ 0.0005 \\
    NEST-mod & $\beta$, $r=0.5319$  & 32.16 $\pm$ 0.03                           & 40.83 $\pm$ 0.03                            & 0.911 $\pm$ 0.006 \\
    \hline
    \hline
    \end{tabular}}
    \caption{The values of $\mathcal{Q}_y$, $\mathcal{L}_y$, and $q$ for the $^{127}$Xe $L$-shell electron capture as measured by XELDA at both 363 and 258~V/cm. Also shown are the values for a 5.2 keV monoenergetic $\beta$ and a 5.2 keV $\beta$ with modified recombination, simulated by NEST with the corresponding recombination probability $r$ as indicated.  Uncertainties in the underlying NEST tritium model are not included in the systematic uncertainties quoted above and should be considered in the application of $\mathcal{Q}_y$ and $\mathcal{L}_y$ as described in Section~\ref{ss:future}.  The unitless ratio $q$ is robust against mismodeling in NEST --- reanalysis with different NEST tritium models gives a 0.3\% shift in $q$ for $^{127}$Xe-L and 0.06\% shifts in $q$ for the simulated $\beta$'s, subdominant to the experimental systematic uncertainty.}
    \label{tab:lshell_yields}
\end{table*}
\par
\begin{figure}[t]
    \centering
    \includegraphics[width=8.6cm, trim=0 0 2.0cm 1.0cm, clip]{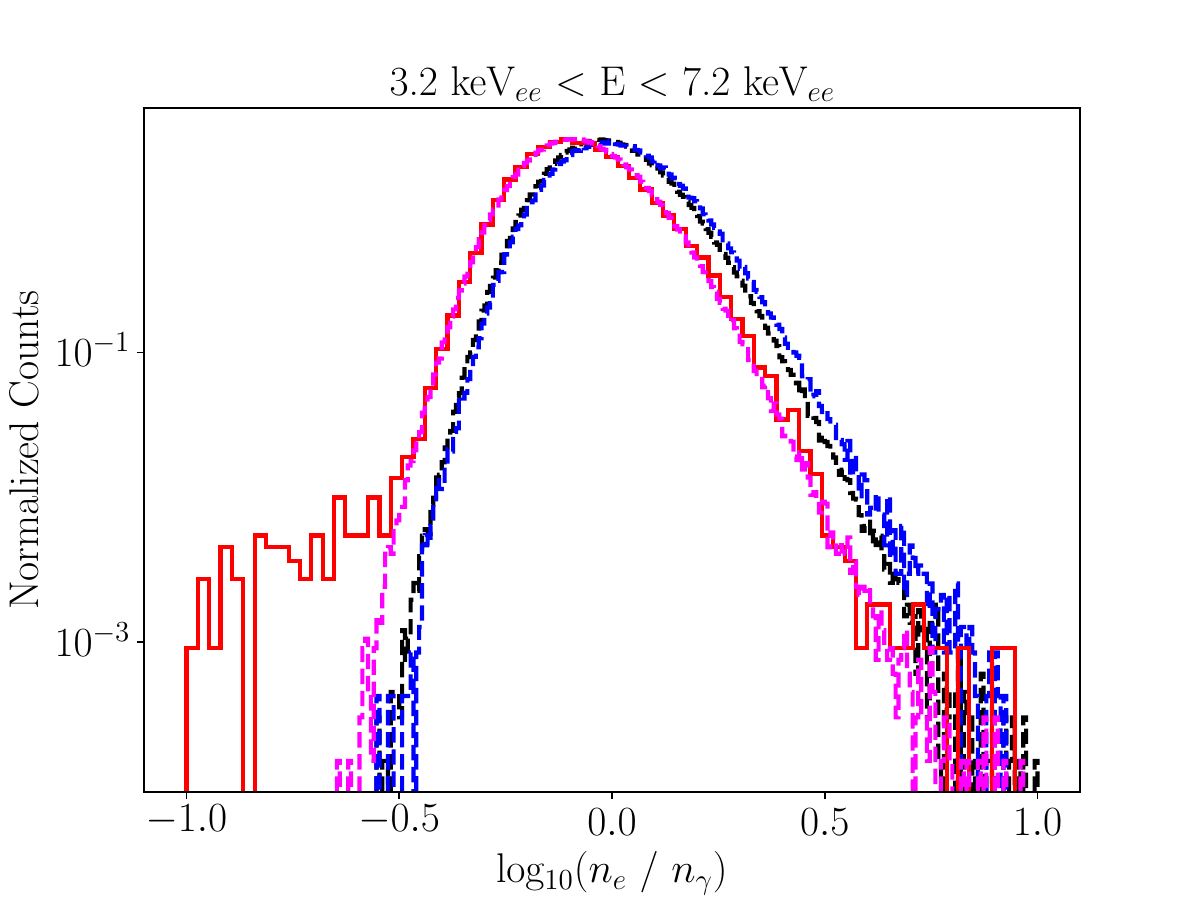}
    \caption{Distributions of discrimination parameters for events within the 3.2--7.2 keV$_{\rm ee}$ region of interest at 363 V/cm. Curves include the $^{127}$Xe $L$-shell as measured by XELDA (red), the NEST expectations for the tritium continuum (blue), a monoenergetic $\beta$ of 5.2 keV (black), and a monoenergetic 5.2 keV $\beta$ with our modified recombination fraction (magenta). The $L$-shell data, as well as our modified recombination beta, show a clear downward shift in $n_e/n_\gamma$ when compared to NEST expectations.}
    \label{fig:logs2s1}
\end{figure}

 Our hypothesis is that the reduced charge yield seen in $L$-shell capture events is due to increased recombination at the event site, and we model this effect by modifying the recombination physics in the NEST package. We float the mean recombination fraction for the simulated mono-energetic ``$\beta$'', leaving the recombination fluctuation model unchanged, and perform a binned maximum-likelihood fit to the observed $\log(n_e/n_\gamma)$ distribution for $L$-shell events. The unmodified and best-fit $\log(n_e/n_\gamma)$ distributions are shown with XELDA data in Fig.~\ref{fig:logs2s1}. The unmodified and best-fit mean recombination fractions ($r$) are listed in Table~\ref{tab:lshell_yields}, along with the simulated charge and light yields using the modified recombination model.  

\section{\label{s:conclusions}Discussion}
\subsection{\label{ss:discrim}Discrimination of $L$-shell capture events}

\par
\begin{figure}[t]
    \centering
    \includegraphics[width=\linewidth]{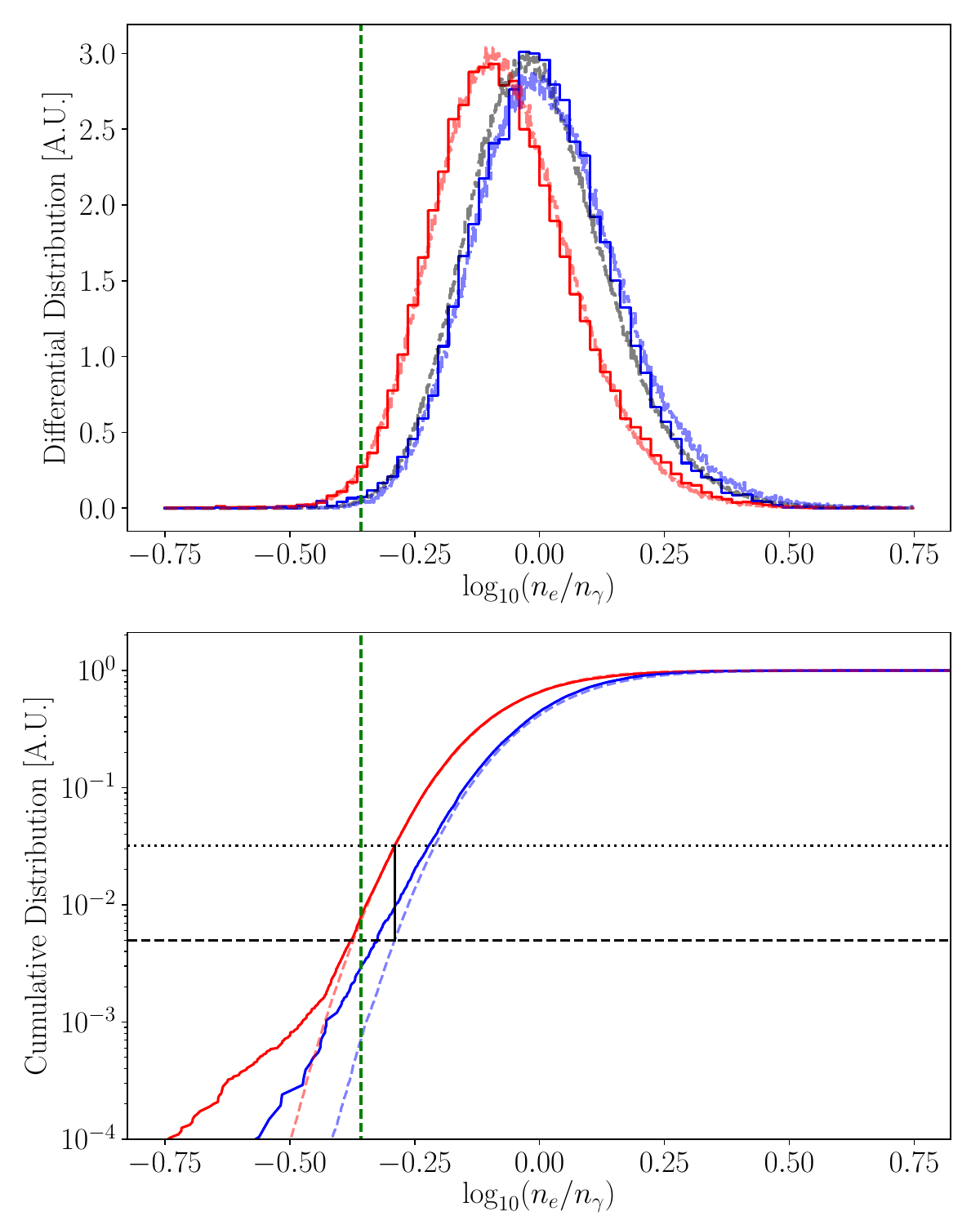}
    \caption{Differential (top) and cumulative (bottom) distributions of $\log_{10}(n_e/n_\gamma)$ for 363 V/cm $L$-shell (solid red) and tritium (solid blue) events in the 3.2--7.2 keV region of interest. The dashed curves show the NEST expectations for a 5.2 keV $\beta$ (black, top-only), the tritium continuum in this region of interest (blue), and the modified-recombination 5.2 keV $\beta$ (red). Also shown is the expected median of the NR distribution (green dashed).  The horizontal dashed and dotted lines in the lower plot show the nominal 0.5\% inefficiency predicted by NEST and 1.65\% inefficiency from the $L$-shell data, respectively, at the cut choice indicated by the vertical black line.  Data diverge from NEST expectations at low acceptance values because the NEST simulation does not include background $\gamma$-X events, wall leakage, and other effects that impact discrimination in small detectors.}
    \label{fig:ANA_logq2lCDF}
\end{figure}

 A key aspect to the success of LXe-TPCs for dark matter is the efficient rejection of ER events using the S2/S1 (or $n_e/n_\gamma$) ratio, and a figure-of-merit for the discrimination power is the fraction of ER events that fall below the NR median. %, nominally $f=0.5$\%, or 1 event in 200. 
We use NEST to predict the NR median in XELDA, shown by the vertical dashed green line in Fig.~\ref{fig:ANA_logq2lCDF}, and to extrapolate the low-$n_e/n_\gamma$ tails of the tritium and $^{127}$Xe $L$-shell spectra, shown by the dashed curves.  The extrapolation is necessary as the observed tails in the XELDA tritium and $L$-shell data are dominated by low-energy external ER backgrounds that are unmodeled in our NEST simulations, and that show anomalously poor discrimination due to detector effects arising in any small chamber, particularly wall effects and multisite events that produce simultaneous energy depositions in the TPC and reverse-field region(s) where the ionization signal is suppressed, also known as $\gamma$-$X$ events.  Our tuned NEST model provides a convenient way to extrapolate spectra to regions that we cannot measure directly due to these backgrounds.
The extrapolated tritium spectrum and NR median give a discrimination power of $7\times10^{-4}$ in XELDA (for the 3.2--7.2 keV window), comparable to that achieved by the LUX collaboration~\cite{ref:LUX_NRrejection},
and an order of magnitude lower than the leakage observed for $L$-shell events.
We note that the modified recombination $\beta$ model accurately reproduces the observed $L$-shell leakage for values of $\log_{10}(n_e/n_\gamma)$ above the NR median.  All of these cases are illustrated for the 363 V/cm data in the bottom panel of Fig.~\ref{fig:ANA_logq2lCDF}. 

\par For a point of comparison where the effect of the unmodeled backgrounds is reduced, we use the value of $\log_{10}(n_e/n_\gamma)$ at which NEST would predict an ER leakage fraction of 0.005 for tritium data. At this value of $\log_{10}(n_e/n_\gamma)$, the $L$-shell leakage fraction for 363 V/cm data is 3.2 times higher than that observed for tritium. Alternatively, a direct comparison to the predicted tritium leakage from NEST without the unmodeled backgrounds finds that the observed $L$-shell rejection inefficiency is a factor of 6.4(6.2) times higher than expected at 363(258) V/cm. In either case, the $L$-shell leakage is significantly worse than what would be expected based on a tritium calibration alone.

\par
\begin{figure}[t]
    \centering
    \begin{subfigure}
        \centering
        \includegraphics[width=\linewidth]{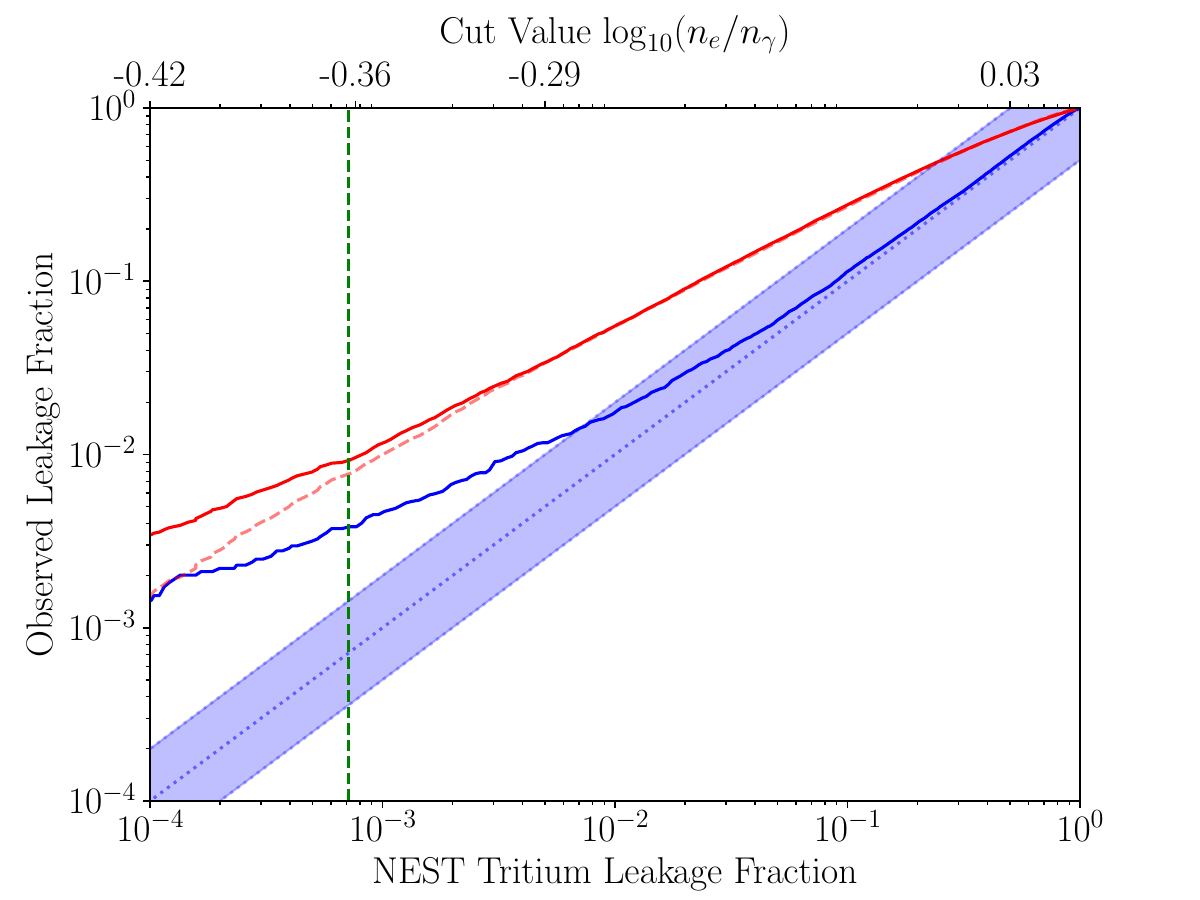}
%        \subcaption{363 V/cm}
    \end{subfigure}
    ~
    \begin{subfigure}
        \centering
        \includegraphics[width=\linewidth]{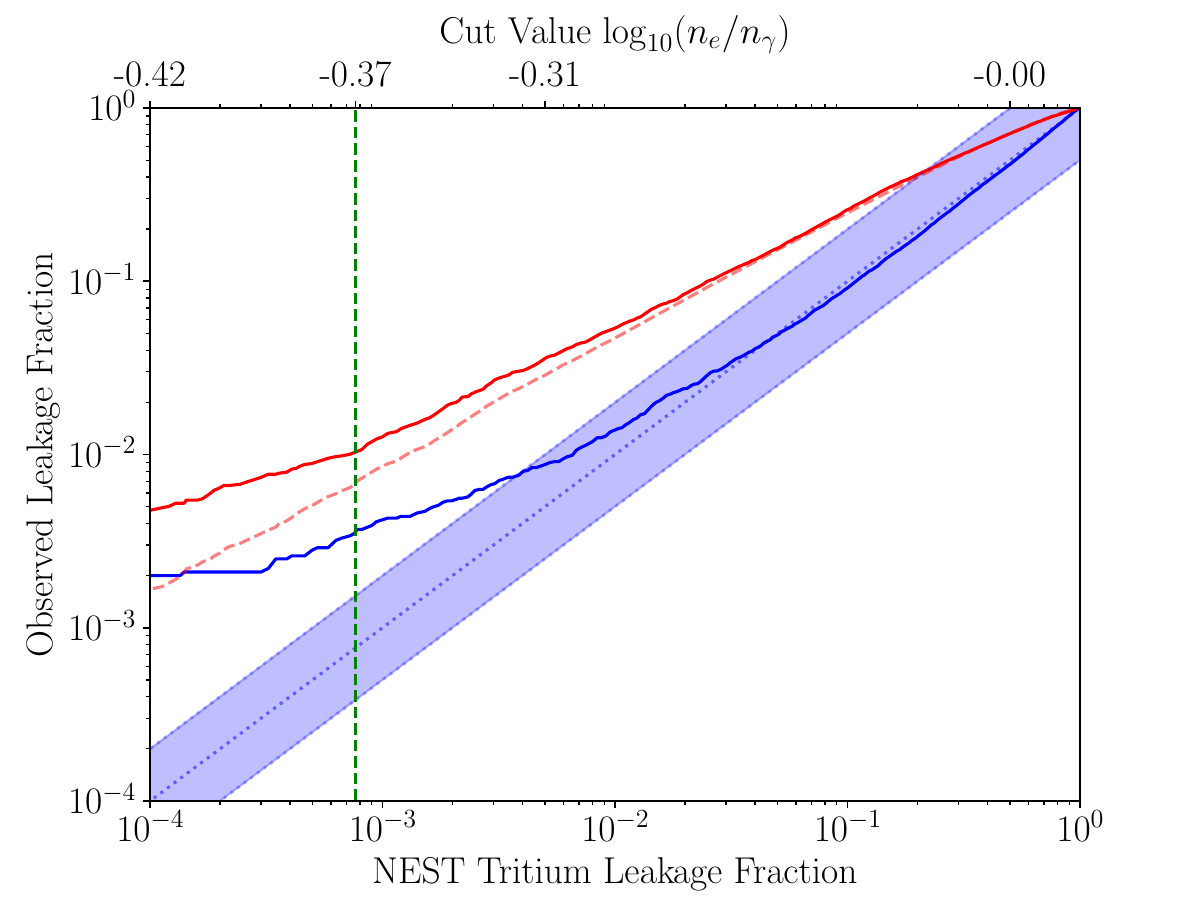}
%        \subcaption{258 V/cm}
    \end{subfigure}
    \caption{The observed leakage fractions of tritium (blue) and $^{127}$Xe $L$ shell (red) for an energy region of 3.2--7.2 keV as a function of NEST's predicted tritium leakage fraction for the 363 (top) and 258 V/cm (bottom) datasets. The blue band shows a factor of 2 around $y=x$ (perfect agreement with NEST), while the red dashed line indicates NEST's simulation of the modified-recombination $\beta$ model. The NR median from NEST is indicated by the vertical green dashed line. The secondary $x$ axis is the value of the cut in $\log_{10}(n_e/n_\gamma)$ that gives rise to the expected and observed leakage fractions.}
    \label{fig:ANA_rejectioneff}
\end{figure}

 For another look at the change in discrimination power, Fig.~\ref{fig:ANA_rejectioneff} shows the observed leakage in both tritium and $L$-shell data as a function of the predicted tritium leakage fraction from NEST, for both drift fields. 
The value of the $\log_{10}(n_e/n_\gamma)$ cut that leads to the predicted NEST tritium leakage rate is shown on the secondary (upper) $x$ axis. 
As in Fig.~\ref{fig:ANA_logq2lCDF}, unmodeled backgrounds induce deviations from NEST predictions in the tails of the observed distributions, but for cut inefficiencies greater than $\sim$1\% the data are robust against the effects of these tails.  We can predict $L$-shell leakage  below 1\% using the modified recombination model, which shows a roughly constant increase in leakage relative to tritium regardless of cut position.  In all scenarios, in both data and simulation, $L$-shell events show significantly greater leakage than is seen for tritium.


\subsection{\label{ss:r+model}Extending the model to $L$-shell scatters}

\par To relate the observed $L$-shell electron capture signal to scatters off $L$-shell electrons, we must model the contribution to the signal of the ejected electron itself.  We expect an $L$-shell scatter event to have the same charge yield as the $^{127}$Xe $L$-shell decay when $E_\mathrm{recoil} =$ 5.2 keV, i.e., when the ejected electron carries no energy. As the energy of the recoil increases, however, more of that energy will be deposited by the ejected electron and the event will approach the same charge yield as a valence scatter. We model this by taking an energy-weighted average of the $L$-shell recombination mean ($r_L$, given in Table~\ref{tab:lshell_yields}) and the recombination mean of a valence shell recoil of energy $E$ ($r_E$, from \cite{ref:NEST} or \cite{ref:LUXTritium}). In this way, the modified inner-shell ER recombination is
\begin{equation}
    r^\prime (E) = \frac{E_L r_{L} + (E-E_L) r_{E}}{E},
\end{equation}
where $E$ is the total deposited energy and $E_L$ is the binding energy of an $L$-shell electron (5.2 keV).  The recombination mean obtained from this average may be less than $r_{E}$, in which case the valence recombination mean should be used instead, giving
\begin{equation}
\label{eq:model}
    r(E) = \left\{\begin{array}{ll}
    r^\prime(E)    &\text{for}~r_E \le r^\prime \\
    r_E      &\text{for}~r_E  > r^\prime
    \end{array}\right..
\end{equation}

\par This model can be applied to the expected solar neutrino ER spectrum~\cite{Vinyoles_2017,ref:LZ_WIMP}, where our corrected recombination factor is applied to $8/52$ of events between $L$- and $K$-shell energies (or to the 8 $L$-shell electrons out of 52 xenon electrons with binding energies in the region of interest).  All other events are generated using the nominal recombination fraction in NEST.
%Such a treatment results in increased leakage for recoils with energy above 5.2 keV.
In the LZ region of interest (1.5--15 keV$_{\rm ee}$), this model predicts a 4.8\% increase in the number of leaked ER events from solar neutrinos over what would be expected from applying the $\beta$ model to the neutrino ER background. % (alternatively, if using an S1 region of interest rather than energy, the leakage increase is only 4.2\%).
The excess events are almost entirely in the 5.2--8.0 keV$_{\rm ee}$ region, where this model predicts a leakage increase of 7.9\%. As NR acceptance goes down this relative increase in overall leakage stays fairly constant, down to at least the 35th percentile of the NR band.

\subsection{\label{ss:neutrino}Impact on dark matter sensitivity}

\par To investigate the impact of this effect on detector sensitivity and dark matter limits, we consider both two-sided profile likelihood ratio (PLR) tests and optimized cut-and-count analyses in three different scenarios, for exposures up to 3~kton-yr.  All three scenarios include backgrounds from neutrino sources only, including ER and NR events from electron- and nuclear-scattering of solar, atmospheric, and supernova neutrinos~\cite{ref:LZ_WIMP,Vinyoles_2017,PhysRevD.90.083510}.  The first two scenarios compare dark matter sensitivity with and without the increased $L$-shell ER leakage, using Eq.~(\ref{eq:model}) and the standard ER recombination model, respectively, to both simulate data and create the background model input to the limit-setting analysis.  The third scenario considers the effect of mismodeling the $L$-shell background, simulating data based on the modified ER recombination model but using the standard ER recombination expectations to build the background model for limit-setting.

\par Both the PLR and cut-and-count analyses show a negligible difference in sensitivity between the first two scenarios, regardless of exposure.  This is not surprising -- while the increase in leakage for $L$-shell capture events on their own is striking, relatively few neutrino events both scatter on the $L$-shell and fall in the narrow energy window where the effect is significant.  The small ($\sim$5\%) increase in overall leakage, when correctly modeled, has virtually no impact on sensitivity, consistent with the analysis performed in Ref.~\cite{pietro_paper}.

\par Using an inconsistent background model when setting a limit has a slightly larger  effect.  In this third scenario, the underestimation of ER leakage leads to a higher limit (overcoverage) than would be obtained using the correct model.  This effect appears in both the PLR and cut-and-count analyses, and the magnitude of the effect grows with exposure --- but remains much smaller than the 1-$\sigma$ experiment-to-experiment variation in the 90\% C.L. upper limit at 50 GeV/$c^2$ dark matter mass for the largest exposure simulated (3 kton-yr).

\par Though this work focuses on the inner-shell effect in the context of the neutrino background, $^{127}$Xe itself is a transient background in LXe-TPC dark matter searches. This background can be modeled directly using the values for observed recombination presented in Sec.~\ref{s:results}.

\subsection{\label{ss:future}Future work}

\par The modified recombination model presented above is sufficient for modeling the relatively small neutrino-electron background expected in the current generation of LXe-TPC dark matter searches, and for planning next-generation searches where neutrino-electron scattering will be the dominant ER background to the weakly interacting massive particle signal. If (or when) these next-generation experiments go forward, further measurements of this effect should be performed to ensure correct statistical coverage and to address questions left unanswered in this work. We have not measured the tails of the S2/S1 distribution associated with $L$-shell vacancies -- instead we assume the same recombination fluctuations that are seen in $\beta$ decays. We have not measured this effect when accompanied by an initially ejected electron -- instead we introduce an \emph{ad hoc} model with reasonable limiting behavior. We have not measured this effect on the $M$ shell, and we have explored only a small range in drift field -- drift-field dependence should be expected, since recombination at higher fields is (in some models) more sensitive to small-scale changes in ionization density. All of these measurements are feasible, many with the $^{127}$Xe technique described here.

\par Recently, Ref.~\cite{ref:xenon_W} published a measurement of the average xenon work function of $W = 11.5$ eV, in agreement with a recent measurement from the EXO-200 collaboration~\cite{ref:exo_W}. This value is $16\%$ lower than the value of 13.7 eV adopted here and in the nominal NEST model, and used widely throughout the field over the last decade. For our analysis, a shift in the value of $W$ would be absorbed by an equal rescaling of the gain factors $g_1$ and $g_2$, and would impact our tritium and xenon-only datasets identically. The values of the absolute quanta yields ($\mathcal{Q}_y$ and $\mathcal{L}_y$) reported in this work would shift by the same amount, but the construction of the ratio $q$ ensures the significance of the $L$-shell effect is unchanged. Similarly, any miscalibration of $g_1$ and $g_2$ stemming from a particular tritium recombination model will propagate into the absolute light and charge yields directly, but the mean $q$ is robust against these effects.  For example, changing from \textsc{NEST} release 2.2.1 patch 1 (used in this analysis) to the current development version (2.3-beta) results in a 5\% decrease (increase) in $L$-shell $\mathcal{Q}_y$ ($\mathcal{L}_y$) due to shifts in the inferred $g_2$ ($g_1$), but $q$ shifts by only 0.3\%.

%when decreasing $g_1$ by $4.2\%$ and increasing $g_2$ by $5.6\%$ both $Q_y$ shift by $5\%$, yet $q$ shifts by $0.3\%$. 

\section{\label{sec:acknowledgements}Acknowledgements}

\par We thank Sidney Cahn, Juan Collar, Carter Hall, Ben Loer, FNAL Proton Assembly Building personnel, FNAL Accelerator Division, and the LZ Collaboration for assistance and advice while performing these measurements. Additionally we would like to acknowledge the contribution from undergraduate and high-school interns: Makayla Trask, Carly KleinStern, Malena Fassnacht, Stella Dang, and Anu Madasi. This work was supported by the Fermi National Accelerator Laboratory, managed and operated by Fermi Research Alliance, LLC, under Contract No. DE-AC02-07CH11359 with the U.S. Department of Energy.  This material is based upon work supported by the U.S. Department of Energy, Office of Science, Office of Workforce Development for Teachers and Scientists, Office of Science Graduate Student Research (SCGSR) program. The SCGSR program is administered by the Oak Ridge Institute for Science and Education for the DOE under contract number DESC0014664. Funding for this work is supported by the U.S. Department of Energy, Office of Science, Office of High Energy Physics under Contracts Number DE-SC0011702 and DE-SC0015910.
%
% \bibliography{main}
% \bibliographystyle{revtex}

\end{document}